\documentclass[twocolumn,prd,aps,showpacs]{revtex4}

\usepackage{amsmath}
\usepackage{verbatim}
\usepackage{graphicx}
\usepackage[usenames]{color}
\usepackage{psfrag}
\usepackage[ps2pdf,colorlinks,bookmarks]{hyperref}

\definecolor{linkblue}{rgb}{0.0,0.0,0.6}
\definecolor{linkgreen}{rgb}{0.0,0.0,0.6}
\hypersetup{pdfpagemode=None, pdfstartview=FitH, linkcolor=linkblue, %
            citecolor=linkgreen, urlcolor=linkblue}

\def\beq{\begin{equation}}
\def\eeq{\end{equation}}
\def\bea{\setlength\arraycolsep{1.4pt}\begin{eqnarray}}
\def\eea{\end{eqnarray}}
\def\bit{\begin{itemize}}
\def\eit{\end{itemize}}

\def\ie{{i.e.}}
\def\eg{{e.g.}}

\def\eq{Eq.~}
\def\eqs{Eqs.~}
\def\fig{Fig.~}

\def\pd{\partial}

\def\ld{\left}
\def\rd{\right}
\def\ra{\rightarrow}
\def\tl{\tilde}
\def\wtl{\widetilde}

\def\fr{\frac}
\def\oo{\frac{1}}

\def\const{{\rm const}}

\def\del{\delta}

\def\Lam{\Lambda}

\def\ad{\dot{a}}

\def\mpl{m_{\rm P}}

\def\O{{\cal O}}
\def\p{{\cal P}}

\def\bra{\langle}
\def\ket{\rangle}

\def\ehat{\boldsymbol{\hat{e}}}
\def\khat{\boldsymbol{\hat{k}}}
\def\alm{a_{\ell m}}
\def\Ylm{Y_{\ell m}}
\def\Cl{C_\ell}
\def\bk{{\boldsymbol{k}}}
\def\bkp{{\boldsymbol{k_\perp}}}
\def\bl{{\boldsymbol{\ell}}}
\def\bx{{\boldsymbol{x}}}
\def\bth{\boldsymbol{\theta}}
\def\R{{\cal R}}
\def\PR{{\cal P_{\cal R}}}
\def\rls{r_{\rm LS}}
\def\xls{x_{\rm LS}}
\def\jl{j_\ell}
\def\camb{\textsc{camb}}

\begin{document}

\title{The Evolution of the Cosmic Microwave Background}

\author{James P.~Zibin} \email{zibin@phas.ubc.ca}
\author{Adam Moss} \email{adammoss@phas.ubc.ca}
\author{Douglas Scott} \email{dscott@phas.ubc.ca}
\affiliation{Department of Physics \& Astronomy\\
University of British Columbia,
Vancouver, BC, V6T 1Z1  Canada}

\date{\today}

\begin{abstract}
   We discuss the time dependence and future of the Cosmic Microwave 
Background (CMB) in the context of the standard cosmological model, in 
which we are now entering a state of endless accelerated expansion.  
The mean temperature will simply decrease until it reaches the effective 
temperature of the de Sitter vacuum, while the dipole will oscillate as 
the Sun orbits the Galaxy.  However, the higher CMB multipoles have a 
richer phenomenology.  The CMB anisotropy power spectrum will for 
the most part simply project to smaller scales, as the comoving distance 
to last scattering increases, and we derive a scaling relation that 
describes this behaviour.  However, there will also be a dramatic 
increase in the integrated Sachs-Wolfe contribution at low multipoles.  
We also discuss the effects of tensor modes and optical depth due to 
Thomson scattering.  We introduce a correlation function relating the 
sky maps at two times and the closely related power spectrum of the 
difference map.  We compute the evolution both analytically and 
numerically, and present simulated future sky maps.
\end{abstract}
\pacs{98.70.Vc, 98.80.Cq, 98.80.Jk}

\maketitle


\section{Introduction}


   The Cosmic Microwave Background (CMB) radiation provides us with a 
vital link to the epoch before the formation of distinct structures,
when fluctuations were 
still linear and carried in a very clean way information about their 
origin, presumably during a phase of inflation.  The simple dynamics of 
the generation and propagation of CMB anisotropies 
(see \eg\ Refs.~\cite{ss06,challinor04,hd02} and references therein) 
depends on a handful of cosmological parameters, $P_i$, such 
as the Hubble constant, the matter density, and spatial curvature, in 
addition to the initial conditions set through inflation.  These dependencies 
have been thoroughly investigated over the past couple of decades and form 
the basis for estimating the parameters from the observed anisotropy spectrum 
of the CMB.

   However, there is one dimension in the parameter space of the CMB 
that has received little explicit attention.  For fixed matter content 
and curvature of the Universe today, we still have the freedom to evolve 
the CMB anisotropies forwards or backwards {\em in time}.  For the practical 
business of performing CMB parameter estimation, 
it is natural of course to suppress this freedom, since we are 
interested in predicting the anisotropies today.  The constraint 
to ``today'' can be applied in at least two ways, which it is 
important to distinguish.  From the set of parameters $P_i$ 
we can calculate the proper-time {\em age\/} of the Universe, $t_0$.  This 
quantity is only determined to an accuracy set by the parameters $P_i$ 
(e.g.\ using Wilkinson Microwave Anisotropy Probe (WMAP) 3-year data, 
Spergel et al. \cite{wmap3} find that 
$t_0 = 13.73^{+0.13}_{-0.17}$ Gyr).  However, the WMAP results 
constrain the redshift of last scattering, $z_{\rm rec}$, (defined as the 
centre of the recombination epoch) to much 
greater accuracy: $z_{\rm rec} = 1088^{+1}_{-2}$ \cite{wmap1params}.  
This very small uncertainty is the result of our accurate determination of 
the mean temperature of the CMB, $T = 2.725\,\pm\,0.001$ K 
\cite{fixsenetal96,mfsmw99}.  
Thus, even though $t_0$ is only known to an accuracy comparable to the 
other parameters $P_i$, implicit in analyses of the CMB is the very tight 
constraint on a different temporal coordinate, $z_{\rm rec}$ or $T$.

   Essentially, the constraint on $t_0$ arises from our determination 
of the expansion rate today, together with information on the content 
and geometry of the Universe, which affect its expansion history.  The 
constraint on $T$ is entirely independent of the content or geometry, hence 
its superior accuracy.  Popular CMB anisotropy numerical packages, such 
as \textsc{cmbfast} or \camb\ \cite{lcl00,sz96} \footnote{Information on 
\camb\ is available at \href{http://camb.info/}{http://camb.info/}.} 
impose the tight constraint arising from the mean temperature through an 
input parameter.  This constraint on $T$ is equivalent, via the 
Stefan-Boltzmann law, to a constraint on the energy density in CMB 
radiation, $\rho_\gamma$.  Therefore by adjusting the temperature input 
parameter, and other cosmological input parameters accordingly, it is 
possible to generate spectra with these packages that correspond to a 
given model evolved into the past or future.  Alternatively it is possible 
to modify the code in these packages to directly integrate to future 
times without the need to modify the input parameters.  Note that if we 
just vary the proper age $t_0$ of a model by varying the expansion rate 
today, this necessarily changes the relative contributions of matter and 
radiation in the past, which affects the physics at recombination and 
hence the shape of the CMB spectrum.

   Fortunately the required code modifications are relatively 
straightforward, and in addition it is possible to describe the temporal 
evolution of the CMB anisotropies analytically to very high precision.  
In this work we systematically describe this evolution both numerically 
and analytically, within the context of the standard $\Lambda$CDM 
($\Lambda$ Cold Dark Matter) model, in order to complete the standard 
results on the parametric dependence of the CMB.  The verification of our 
numerical work with our analytical results, and conversely the 
characterization of our analytical approximations with the full numerical 
calculations, will be crucial in this novel study.  We will find that 
while the temporal behaviour of the CMB power spectrum is 
determined mainly by a simple geometrical scaling relationship, less 
trivial physics arises when we consider the behaviour of {\em correlations\/} 
between anisotropies at different times.

   It can certainly be argued that the standard calculations of the CMB 
anisotropy spectrum implicity describe its time dependence in that the 
spectrum must be evolved from the time of recombination to the present.  
Nevertheless, there appear to have been very few explicit discussions 
of the time dependence, with the exceptions being primarily concerned with 
the distant future.  Gott \cite{gott96} points out that at extremely 
late times the typical wavelengths of CMB radiation will exceed the 
Hubble radius, and so the CMB radiation will be lost in a de Sitter 
background.  Loeb \cite{loeb02} mentions that as the time of 
observation increases, the radius of the last scattering sphere also 
increases, and approaches a maximum in a $\Lambda$CDM model.  This leads 
to the potential for reducing the cosmic variance limitation on the 
determination of the anisotropy spectrum.  Krauss and Scherrer \cite{ks07} 
point out that well before this final stage, the CMB will redshift below the 
plasma frequency of the interstellar medium and hence be screened from 
view inside galaxies.

   Importantly, when discussing the distant future evolution of the Universe 
it must be remembered that even the {\em qualitative\/} details can depend 
very sensitively on the model adopted.  An extreme example is the potential 
destruction of the Universe in finite proper time in a ``big rip'' 
\cite{ckw03}, 
when the dark energy violates the weak energy condition, with equation of 
state $w < -1$.  In the present work, for the sake of definiteness and 
simplicity, we conservatively choose a spatially flat model in which 
the dark energy is a pure cosmological constant, with cosmological parameters 
consistent with the WMAP results \cite{wmap3}.  However, using the 
techniques we discuss it is straightforward to extend our results to 
other specific dark energy models.

   An interesting question that naturally arises in the present context is: 
How long must we wait before we could observe a change in the CMB?  The 
formalism that we develop here will be necessary to answer this 
question, and we will address this explicit issue in separate work 
\cite{mzs07}.  We stress that the practicalities of an 
experimental search for the effects we describe is not our concern in the 
present work.  In addition to setting the stage for the detectability 
question, we believe that a discussion of the evolution of the 
CMB sky is useful in its own right in elucidating the physics of the 
anisotropies in a vacuum-dominated cosmology.

   We begin in Section~\ref{sec:bulk} with a description of the time 
dependence of the ``bulk'' properties of the CMB, namely the mean 
temperature and the dipole.  After a brief review of the formalism used 
to describe the anisotropy power spectrum, its evolution is described 
analytically in Section~\ref{sec:ps}, including the effects of the 
integrated Sachs-Wolfe effect, tensor modes, and reionization, and numerical 
calculations are presented using our modified version of the line-of-sight 
Boltzmann code \camb.  In Section~\ref{sec:diffmap} 
we introduce the difference map power spectrum and associated correlation 
function, and present analytical and numerical calculations.  
Section~\ref{sec:discuss} presents our conclusions, and in the \hyperref
[flatskyapp]{Appendix} a description of an important approximation method 
is presented.  We set $c = 1$ throughout.


\section{Time evolution of the bulk CMB}
\label{sec:bulk}

   The temperature fluctuations on the CMB sky can be decomposed into a 
set of amplitudes of spherical harmonics (see Section~\ref{sec:form}).  
The angular mean temperature (or ``monopole'') and dipole have a special 
status.  The mean temperature is just a measure of the local radiation 
energy density, while the value of the dipole depends on the observer's 
reference frame at linear order (higher multipoles are independent of 
frame at this order).

\subsection{The mean temperature}

   As time passes, the change in the CMB that is simplest to quantify is 
the cooling of its mean temperature $T$ due to the Universe's expansion.  
The CMB radiation was released from the matter at the time of last 
scattering, when $T \simeq 3000$ K.  It later reached a comfortable $300$ K 
at an age of about $t\simeq 15$ Myr, and is now only a frigid few Kelvin.  
Indeed, the monotonicity of the function $T(t)$ means that $T$ {\em itself\/} 
can be used as a good time variable.  Thus we can consider measurements 
of $T$ as direct readings of a sort of ``cosmic clock''.

   Today the CMB radiation is essentially free streaming, \ie\ 
non-interacting with the other components of the Universe.  Therefore the 
energy density in the CMB evolves according to the energy 
conservation equation $\dot\rho_\gamma = -4H\rho_\gamma$, where $H$ is 
the Hubble parameter and the overdot represents the proper time 
derivative.  Since $\rho_\gamma \propto T^4$, we have $\dot T = -HT$.  
Evaluating this expression today, using $T_0 = 2.725$ K and $H_0 = 73$ 
km$\,$s$^{-1}\,$Mpc$^{-1}$ (subscript $_0$ indicates values today), we find
\beq
\dot T_0 = -0.20\ \mu\mathrm{K}\ \mathrm{kyr}^{-1}.
\eeq
Thus in $5000$ yr the mean temperature will drop by $1\ \mu$K.

   The CMB radiation continues to redshift indefinitely as the Universe 
expands in the late $\Lambda$-dominated de Sitter phase.  However, this 
does not mean that as the CMB becomes increasingly difficult to 
measure, clever experimentalists need only to ever refine their instruments 
in order to keep up.  Instead, a fundamental limit exists below which 
a CMB temperature cannot be sensibly defined.  An object in an otherwise 
empty de~Sitter phase will see a thermal field with temperature \cite{gh77}
\beq
T_{\rm dS} = \fr{H}{2\pi} = \oo{2\pi}\sqrt{\fr{\Lambda}{3}}.
\eeq
Therefore, after the CMB temperature redshifts to below 
$T_{\rm dS}$, the CMB becomes lost in the thermal noise of the 
de~Sitter background, as pointed out in \cite{gott96} (see also 
\cite{bawe03}).

   To see explicitly the difficulty with measuring the CMB at such 
extremely late times, consider the typical wavelengths 
of radiation in the CMB.  A thermal spectrum at temperature $T$ consists 
of wavelengths $\lambda$ on the characteristic scale $T^{-1}$ (the precise 
peak position of the Planck spectrum depends on the measure used for the 
thermal distribution).  Therefore when $T = T_{\rm dS}$ we have 
$\lambda \simeq H^{-1}$, \ie\ the typical CMB wavelengths become of order 
the Hubble length.  Alternatively, at late times in the 
de~Sitter phase the frequency of a mode of CMB radiation of fixed comoving 
wavelength redshifts according to
\beq
\omega(t) = \omega(t_1) e^{-H_{\rm dS}(t - t_1)},
\eeq
where
\beq
H_{\rm dS} \equiv \sqrt{\Omega_\Lambda} H_0
\eeq
is the asymptotic value of the Hubble parameter and $t_1$ is some late 
proper time.  Therefore the accumulated phase shift between time $t_1$ and 
the infinite future is
\beq
\int_{t_1}^\infty \omega(t)dt = \fr{\omega(t_1)}{H_{\rm dS}}.
\eeq
This expression tells us that when the frequency becomes less than the 
Hubble parameter (\ie\ the wavelength becomes larger than the Hubble length), 
a full temporal oscillation cannot be observed, even if we observe into 
the infinite future.  In terms of conformal time, the oscillation rate 
remains constant in the de~Sitter phase, but there is only a finite amount 
of conformal time available in the future.  Indeed, considering the quantum 
nature of such a mode, this calculation provides insight into the necessity 
of a residual de~Sitter thermal spectrum at this scale.

   The energy density in the CMB at arbitrary scale factor $a$ is given by
\beq
\rho_\gamma = \fr{3H_0^2\mpl^2\Omega_\gamma}{8\pi}\ld(\fr{a_0}{a}\rd)^4,
\eeq
where $\mpl$ is the Planck mass and $\Omega_\gamma = 5 \times 10^{-5}$ 
is the fraction of the total density in the CMB today.  Using this 
expression we can show that we 
must wait until $a/a_0 \sim 10^{30}$ before $\rho_\gamma 
= \rho_{\rm dS} \equiv T_{\rm dS}^4$ and the CMB becomes lost in the 
de~Sitter background.  This corresponds to an age of $t = 1$~Tyr.  
If we ask instead at what scale factor would the 
radiation density be equal to the Planck density, $\rho_{\rm P} = 
\mpl^4$, we find $a/a_0 \sim 10^{-32}$.  It might appear, 
therefore, that we exist at a special time, in that the radiation density 
today is roughly $120$ decades removed from both the Planck era and the 
final era when $T = T_{\rm dS}$.  To understand the origin of this 
coincidence, note that by virtue of the above expressions and the energy 
constraint (or Friedmann) equation, the 
three densities $\rho_{\rm P}$, $\rho_\Lambda$, and $\rho_{\rm dS}$ are 
in the geometrical ratio $\mpl^4:\Lambda\mpl^2:\Lambda^2$, 
up to numerical factors.  Therefore, the apparent coincidence 
just described is actually equivalent to the standard coincidence problem, 
namely that $\rho_\Lambda \simeq \rho_{\rm tot}$ today, given that 
$\rho_\gamma$ differs from $\rho_{\rm tot}$ today by ``only'' a few 
decades.  {\em Any\/} density that today even crudely approximates the dark 
energy density will necessarily be separated by roughly $120$ decades from 
both $\rho_{\rm P}$ and $\rho_{\rm dS}$.

\subsection{The dipole}

   The observed dipole anisotropy in the CMB can be attributed to the 
Doppler effect arising from our peculiar velocity, $\boldsymbol{v}$, with 
respect to the frame in which the CMB dipole vanishes.  That peculiar 
velocity, and hence the dipole, is expected to evolve with time.  The 
magnitude of the dipole can be specified by the maximum CMB temperature 
shift over the sky, $\delta T_d$, due to the velocity 
$\boldsymbol{v}$.  This is given by the lowest order Doppler expression,
\beq
\fr{\delta T_d}{T} = v.
\label{doplerlin}
\eeq
(In terms of the spherical harmonic expansion to be introduced 
in \eq(\ref{texp}), we have $\delta T_d/T = \sqrt{3/(4\pi)}a_{10}$, when the 
polar axis is aligned with $\boldsymbol{v}$.)

   The current best estimate of the magnitude of the dipole comes from 
observations of the WMAP satellite---indeed, the annual 
modulation by the Earth's motion 
around the Sun is actually used to calibrate satellite experiments, so 
this aspect of the time-varying dipole is already well determined.  The 
measured value of the dipole, in Galactic polar coordinates, is 
$(\delta T_d,l,b)=(3.358 \pm 0.0017$ mK, $263.86 \pm 0.04^\circ$, 
$48.24 \pm 0.10^\circ)$ \cite{hinshaw06}.  Equivalently, the 
Cartesian velocity vector is 
$\boldsymbol{v}_0 \simeq (-26.3$, $-244.6$, $275.6)$ km$\,{\rm s}^{-1}$, 
where the first component is towards the Galactic centre, and the 
third component is normal to the Galactic plane.  
Therefore, in natural units we have $v_0 = 1.2 \times 10^{-3}$, so 
the lowest order approximation, \eq(\ref{doplerlin}), is valid.

   In order to determine the evolution of the dipole, we could 
straightforwardly calculate it at linear order.  However, linear theory is 
certainly not a good approximation on sufficiently small scales today.  To 
fully describe the evolution of the velocity $\boldsymbol{v}$ we must take 
into account the presence of the nonlinear structures we observe on small 
scales today.

   This velocity vector can be considered as a sum of individual vectors 
contributing to the overall motion of the Sun with respect to the CMB.  In 
the local neighbourhood, the Sun moves with respect to the ``local standard 
of rest'', which in turn moves with respect to the Galactic centre.  However, 
the peculiar motions in the Solar neighbourhood are at the 10\% level compared 
with the motion of the Sun around the Milky Way \cite{kogut93}, so for 
the purposes of the simple calculation which follows we ignore these 
contributions.  Also, we will consider here time scales short enough 
that the motion of the Milky Way within the Local Group, and the Local 
Group relative to Virgo, the Great Attractor, and other distant 
cosmic structures is approximately constant (see, \eg, \cite{tully07} for 
a description of these motions).

   Just as today we can detect the modulation of the Earth's motion around 
the Sun, in the future, with increasing satellite sensitivity, we may be 
able to observe the Sun's motion around the Galaxy.  For the motion of the 
Sun around the Milky Way, we assume that this is simply a tangential speed 
of $220$ km$\,{\rm s}^{-1}$ at a distance of $8.5$ kpc.  Using the current 
observed value of $\boldsymbol{v}$ to infer the velocity of the Galactic 
centre with respect to the CMB rest frame, the time dependent Sun-CMB 
velocity vector is then
\bea
\boldsymbol{v}(t)=\bigg[\!\!\!&&222\sin\ld(\frac{2\pi t}{T}\rd)
 -26.3,\nonumber\\
 &&222\cos\ld(\frac{2\pi t}{T}\rd) - 466.6,\:275.6\bigg]
 {\rm km}\,{\rm s}^{-1}\,,
\label{eqn:suncmb}
\eea
where the Galactic orbital period is $T=2.35 \times 10^{8}$ yr.

   In order to ascertain when a change in the dipole is detectable, one 
could compute a sky map of the dipole at two times.  If the temperature 
variance of the difference map is greater than the experimental noise 
variance, then a detection is probable.  In this case, 
the variance of the difference map, which we denote by $C_S$, 
is $C_S=[(\delta v_{x})^{2}+ (\delta v_{y})^{2}+ 
(\delta v_{z})^{2}]/(4 \pi)$, where $\del \boldsymbol{v}$ is the 
difference of the Sun-CMB dipole vector between the two observations.  
Using \eq(\ref{eqn:suncmb}), and converting to fractional temperature 
variations, the signal variance of the changing dipole is then
\begin{equation} 
C_S=8.7 \times 10^{-8} \left[ 1- \cos \left( \frac{2 \pi t}{T} 
\right) \right]\,.
\end{equation}
Later in this paper we compute signal variances involving 
higher order CMB multipoles.  These variances are of course much smaller 
than that of the dipole.  In a follow up paper 
\cite{mzs07} we will discuss in detail the prospects for detecting a 
change in the CMB with future experiments.

   Finally, we note that in this simple calculation it is assumed that we 
have a frame of reference external to the Milky Way in order to 
construct our coordinate system. This could be provided, for example, 
by the International Celestial Reference Frame, based on the 
positions of 212 extra-galactic sources \cite{ma98}.


\section{The anisotropy power spectrum}
\label{sec:ps}

\subsection{Review of the basic formalism}
\label{sec:form}

   There is much more information encoded in the anisotropies of the CMB 
than in the mean temperature, since the anisotropies are determined by the 
details of the matter and metric fluctuations near the last-scattering 
surface (LSS) and all along our past light cone to today.  Therefore it 
is much less trivial to determine the time evolution of the anisotropies 
than the mean temperature (or dipole).  However, in the approximation that all 
of the CMB radiation was emitted from the LSS at some instant $t_{\rm LS}$ 
when electrons and photons decoupled, and then propagated freely, the 
evolution of the primary power spectrum of the CMB is determined by a 
simple geometrical scaling relation which is closely related to 
the main geometrical parameter degeneracy in CMB spectra.  In order to 
derive this relation, and to describe the behaviour of the correlation 
functions introduced in later sections, it will be helpful to first 
summarize the standard description of CMB anisotropies in a form that 
will be easy to generalize.  This subsection may be skipped by readers 
familiar with the material.  For detailed treatments of the generation of 
anisotropies see \eg\ \cite{hs95,dodelson03}.

   At very early times, when each perturbation mode, labelled by comoving 
wavevector $\bk$, is 
outside of the Hubble radius, the fluctuations can be described by a single 
perturbation function, for the case of adiabatic perturbations.  It is 
very convenient to take this function to be the curvature perturbation on 
comoving hypersurfaces, $\R$, since this quantity is conserved on large 
scales in this case, and hence can be readily tied to the predictions of 
a specific inflationary model.  In the simplest models of inflation, $\R$ 
is predicted to be a Gaussian random field to very good approximation, 
fully described by the relation
\beq
\bra\R^\ast(\bk)\R(\bk')\ket = 2\pi^2\delta^3(\bk - \bk')\fr{\PR(k)}{k^3},
\label{primps}
\eeq
with primordial power spectrum $\PR(k)$ and $k \equiv |\bk|$.  For a 
scale-invariant spectrum we have $\PR(k) =$ constant.

   The fluctuations at last scattering can be described by a set of matter 
and metric perturbations, $\phi_i(\bx,\eta)$, where for future convenience we 
have used comoving coordinate $\bx$ and conformal time $\eta$.  Since 
linear perturbation theory is a very good approximation at the scales 
sampled by the CMB, this set of perturbations is determined from the 
primordial comoving curvature perturbation by transfer functions 
$A_i(k,\eta)$ via
\beq
\phi_i(\bk,\eta) = A_i(k,\eta)\R(\bk).
\label{phitransfr}
\eeq
In the approximation of abrupt recombination, so that the LSS has zero 
thickness, followed by free streaming of radiation, the observed primary 
temperature anisotropy $\delta T(\ehat)/T$ in direction $\ehat$ is 
determined by the fluctuations at the corresponding point on the LSS, \ie
\beq
\fr{\delta T(\ehat,\eta)}{T(\eta)} = F(\phi_i(\rls,\ehat,\eta_{\rm LS})),
\label{sw1}
\eeq
for some linear function $F$.  Here $\rls = \eta - \eta_{\rm LS}$ is the 
comoving radial coordinate to the LSS from the point of observation, taken 
to be the origin of spherical coordinates.  \eq(\ref{sw1}) ignores both 
the effect of gravitational lensing by foreground structure and the 
effect of reionization at late times (we examine the effects of 
reionization in Section~\ref{sec:optdepth}).  In the approximation that 
photons are tightly coupled to baryons before $\eta_{\rm LS}$, the function 
$F$ can be written in terms of two perturbation functions as
\beq
F(\phi_i(\rls,\ehat,\eta_{\rm LS})) = \phi_1(\rls,\ehat,\eta_{\rm LS}) + 
                     \fr{\pd}{\pd\rls}\phi_2(\rls,\ehat,\eta_{\rm LS}).
\label{sw2}
\eeq
\eqs(\ref{sw1}) and (\ref{sw2}) describe the generation of CMB anisotropies 
through the Sachs-Wolfe effect \cite{sw67}, with the first term on the 
right-hand side 
of (\ref{sw2}) the so-called ``monopole'' contribution, and the second term 
the ``dipole'' or ``Doppler'' contribution.

\begin{figure}[ht]
\includegraphics[width=0.9\columnwidth]{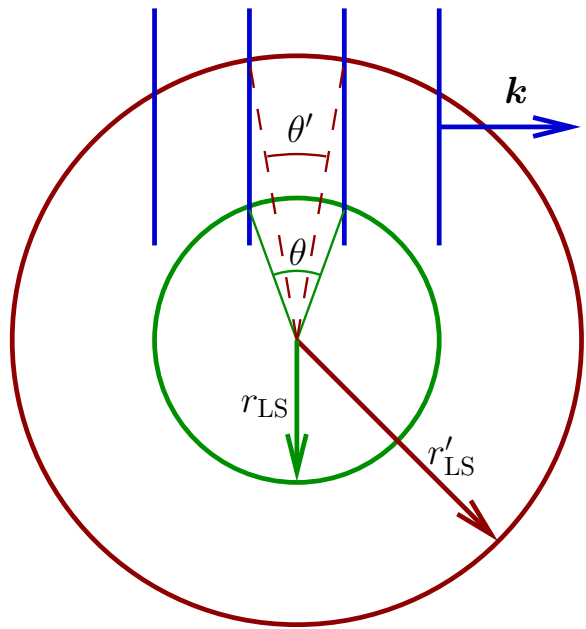}
\caption{Our spheres of last scattering at time $\eta$ (inner) and $\eta'$ 
(outer).  The group of vertical lines indicates the crests of a mode $\bk$, 
and hot spots will be observed at the intersections of those crests with 
the spheres.  One wavelength of the mode will span angle $\theta$ at $\eta$ 
but a smaller angle $\theta'$ at $\eta'$.}
\label{lssfig}
\end{figure}

   The preceeding equations have the very simple interpretation that when 
we measure the CMB anisotropies at some time $\eta$ we are ``seeing'' the 
primordial fluctuations $\R$ on the comoving spherical shell $r = \rls = 
\eta - \eta_{\rm LS}$, as processed by the linear transfer functions $A_1$ 
and $A_2$ to the time $\eta_{\rm LS}$.  If we observe at a later time 
$\eta'$, we see the fluctuations on a larger shell of radius $r = \rls' 
\equiv \eta' - \eta_{\rm LS}$, as illustrated 
in \fig\ref{lssfig}.  The fluctuations at the LSS contain structure 
at various scales, encoded in the transfer functions, due to acoustic 
oscillations within the pre-recombination plasma.  Assuming the statistical 
homogeneity of space, that structure will occur at the same physical 
scales on the shells $r = \rls$ and $r = \rls'$.  Therefore we expect that 
structure visible at time $\eta$ on angular scale $\theta$ will also be 
visible at $\eta'$, but at the smaller angular scale
\beq
\theta' = \theta\,\fr{\rls}{\rls'},
\label{thetascal}
\eeq
at least for small scale structure, $\theta \ll 1$.  To make this rigorous, 
and to derive in addition the scaling law for the {\em amplitude\/} of angular 
structure, we need to next introduce a spherical expansion of the CMB 
anisotropy.  

   We expand as usual the temperature fluctuation observed in some 
direction $\ehat$ in terms of spherical harmonics $\Ylm$ as
\beq
\fr{\delta T(\ehat,\eta)}{T(\eta)} = \sum_{\ell m}\alm(\eta)\Ylm(\ehat).
\label{texp}
\eeq
The expansion coefficients $\alm$ determine all the details of the 
particular sky map of the CMB observed at time $\eta$.  However, the 
statistical properties of the $\alm$s are determined through 
\eqs(\ref{phitransfr}) to (\ref{sw2}) by the statistics of $\R$ encoded in 
\eq(\ref{primps}).  To make this explicit, we need the spherical expansion 
of the perturbations $\phi_i(\rls,\ehat,\eta_{\rm LS})$, namely
\beq
\phi_i(\rls,\ehat) = \sqrt{\fr{2}{\pi}} \int kdk \sum_{\ell m} 
                     \phi_{i\ell m}(k) \jl(k\rls)\Ylm(\ehat).
\label{phiexp}
\eeq
Here $i = 1$ or $2$, $\jl$ is the spherical Bessel function of the first kind, 
and we have dropped the understood argument $\eta_{\rm LS}$.  Next the identity
\beq
f_{\ell m}(k) = ki^\ell\int d\Omega_\bk f(\bk)\Ylm(\khat)
\eeq
(see, \eg, Ref. \cite{ll00}) combined with \eq(\ref{phitransfr}) allows us to 
write
\beq
\phi_{i\ell m}(k) = A_i(k)\R_{\ell m}(k).
\label{phirsph}
\eeq
Now, combining \eqs(\ref{sw2}), (\ref{phiexp}), and (\ref{phirsph}), we have
\beq
F(\rls,\ehat) = \sqrt{\fr{2}{\pi}} \int kdk \sum_{\ell m} \R_{\ell m}(k)
   T(k,\ell,\rls)\Ylm(\ehat),
\label{frtransf}
\eeq
where
\beq
T(k,\ell,\rls) \equiv A_1(k)\jl(k\rls) + A_2(k)\jl'(k\rls)
\label{tdef}
\eeq
and the prime denotes differentiation with resepect to $\rls$.  Finally, 
equating coefficients between \eq(\ref{frtransf}) [with \eq(\ref{sw1})] 
and \eq(\ref{texp}), we obtain
\beq
\alm(\eta) = \sqrt{\fr{2}{\pi}} \int kdk \R_{\ell m}(k) T(k,\ell,\rls),
\label{almrtr}
\eeq
where we have restored the argument $\eta = \rls + \eta_{\rm LS}$.  This 
expression gives the CMB anisotropy in terms of the primordial 
perturbations $\R$ and a new linear transfer function $T(k,\ell,\rls)$.

   In order to determine the statistical properties of the $\alm$s, 
we need the expression
\beq
\bra\R_{\ell m}(k) \R^\ast_{\ell' m'}(k')\ket = 
   2\pi^2\delta(k - k')\fr{\PR(k)}{k^3}\delta_{\ell\ell'}\delta_{mm'},
\eeq
which can be derived from \eq(\ref{primps}).  Using this expression and 
\eq(\ref{almrtr}) we find
\beq
\bra\alm(\eta) a^\ast_{\ell'm'}(\eta)\ket = \Cl(\eta)\delta_{\ell\ell'}
\delta_{mm'},
\label{almcl}
\eeq
where
\beq
\Cl(\eta) \equiv 4\pi\int\fr{dk}{k}\PR(k)T^2(k,\ell,\rls).
\label{cldef}
\eeq
That is, each coefficient $\alm$ has variance $\Cl(\eta)$ (which is called the 
anisotropy power spectrum) and coefficients for different spherical modes 
are uncorrelated.

   Note that \eq(\ref{almrtr}), and hence \eqs(\ref{almcl}) and 
(\ref{cldef}), hold even when 
we relax the tight coupling and free streaming approximations, with {\em 
some\/} transfer function $T(k,\ell,\rls)$.  However, in the general case 
the transfer function must be calculated numerically.


\subsection{Analytical time evolution for primary anisotropies}
\label{sec:analps}

   The formalism developed in the preceeding subsection can now be applied 
to describe the time evolution of the primary CMB anisotropy spectrum, under 
the abrupt recombination and free streaming approximations.  To determine 
the time evolution of $\Cl(\eta)$, \eq(\ref{cldef}) tells us that we 
only need to consider the behaviour of $T^2(k,\ell,\rls)$ as $\rls$ 
increases (note that we will often adopt the coordinate $\rls$ as an 
effective time coordinate).  To do this, \eq(\ref{tdef}) tells us that we 
only require the behaviour of the products $\jl^2(k\rls)$, $\jl'^2(k\rls)$, 
and $\jl(k\rls)\jl'(k\rls)$ as functions of $\rls$.  This can be done 
in the limit $\ell \gg 1$ using asymptotic forms for the Bessel functions.  
For large $\ell$ we can write [see Ref. \cite{as72}, \eq(9.3.3)]
\beq
j_\ell(x) = (x^4 - x^2\ell^2)^{-1/4}[\cos(\theta) + \O(1/\ell)],
\quad \textrm{for }x > \ell,
\eeq
where $\theta = \theta(x)$ is a real phase.  For $x < \ell$, $j_\ell(x)$ 
decays rapidly.  This allows us to write a scaling relation for the envelope 
of the Bessel oscillations, namely
\beq
j_\ell(x) \sim \alpha j_{\alpha\ell}(\alpha x),
\label{besscal}
\eeq
for positive $\alpha$ such that $\alpha\ell \gg 1$ also applies.  This 
expression will allow us to obtain the time dependence of the ``monopole'' 
contribution to $T^2(k,\ell,\rls)$, which is proportional to 
$\jl^2(k\rls)$.  (The Bessel oscillations are rapid relative to the range of 
scales that contribute to the integral \eq(\ref{cldef}) for $\ell \gg 1$, and 
hence can be ignored.)  We can write the ``dipole'' part $\jl'^2(k\rls)$ in 
terms of spherical Bessel functions using recurrence relations and again 
apply \eq(\ref{besscal}) to obtain the time dependence.  The cross term 
proportional to $\jl(k\rls)\jl'(k\rls)$ can be shown to be negligible, \ie\ 
the monopole and dipole contributions add incoherently.  Applying 
\eq(\ref{besscal}), then, we find that for large $\ell$,
\beq
T^2(k,\ell',\rls') \simeq \fr{\rls^2}{\rls'^2}T^2(k,\ell,\rls),
\label{transscal}
\eeq
where we have defined
\beq
\ell' = \ell\fr{\rls'}{\rls}.
\label{lscal}
\eeq
Applying \eq(\ref{cldef}) we finally obtain the scaling relation for the 
power spectrum,
\beq
\ell'^2C_{\ell'}(\eta') \simeq \ell^2C_{\ell}(\eta).
\label{clscal}
\eeq
Importantly, \eq(\ref{clscal}) holds independently of the form of the 
functions $A_i(k)$ and $\PR(k)$, so the result applies to the acoustic peak 
structure as well as to non-scale-invariant primordial spectra.

   This result confirms our previous prediction, \eq(\ref{thetascal}), for 
the dependence of angular scales on observation time.  But the dependence 
of the {\em amplitude\/} of the spectrum encoded in \eq(\ref{clscal}) is 
also not surprizing, since the quantity $\ell(\ell + 1)C_{\ell}$ 
is independent of $\ell$ in the Sachs-Wolfe plateau for a scale invariant 
spectrum, as is well known.  But the height of that plateau, calculated 
using $A_1 = \const$ and $A_2 = 0$ above, is independent of the observation 
time.  (Indeed that height is, up to numerical factors, simply $\PR$.  Recall 
that $\Cl$ is determined by the ratio $\del T/T$.  The absolute anisotropies 
$\del T$ exhibit the same expansion redshift as does the mean temperature 
$T$.) Hence as $\eta$ increases, the quantity $\ell^2C_{\ell}(\eta)$ must 
remain constant (up to corrections of order $1/\ell$), which is precisely 
what \eq(\ref{clscal}) says.  Of course, the result (\ref{clscal}) is valid 
for the entire acoustic peak structure, not just the Sachs-Wolfe plateau.

   The result (\ref{clscal}) is derived in the 
\hyperref[flatskyapp]{Appendix} much more directly, without resorting to 
properties of Bessel functions, using the flat sky approximation.  In that 
approach 
we consider anisotropies in a patch of sky small enough that it can be 
approximated as flat, and errors are again of order $1/\ell$.

   In addition to the main temperature anisotropies we have been considering 
here, there are also polarization spectra present in the CMB radiation.  The 
polarization is sourced primarily near last scattering, so its spectra will 
also scale according to \eq(\ref{clscal}).  A small part of the largest-scale 
polarization is sourced near reionization, so we expect that that 
contribution will scale with the comoving radius to the reionization 
redshift, rather than to the last scattering surface.

   Having found the scaling relation (\ref{clscal}), we can next derive 
some simple consequences from it.  First, we can write the total power in 
the anisotropy spectrum as
\beq
\sum_{\ell m} \Cl(\eta) = \sum_\ell (2\ell + 1) \Cl(\eta)
                        \simeq 2\int \ell \Cl(\eta) d\ell
\eeq
in the large $\ell$ approximation.  Then, using \eq(\ref{clscal}), we have
\beq
\sum_\ell (2\ell + 1) \Cl(\eta) \simeq \sum_\ell (2\ell + 1) \Cl(\eta'),
\eeq
where the approximation comes from ignoring terms of order $1/\ell$.  What 
this expression says is that 
the total power is constant in time, for the free streaming of primary 
anisotropies.  This result is equivalent to the ``conservation condition'' 
stated in \cite{be87}.  Implicit in this result is the assumption 
of statistical homogeneity, so that no new anomalous power will be 
revealed at the largest scales as $\rls$ increases.  As we will see 
in Section~\ref{sec:isw} below, secondary anisotropies, in particular those 
generated through the ``integrated Sachs-Wolfe effect'', are expected to grow 
dramatically at late times and hence the total power will not in fact 
be conserved.

   Another consequence of \eq(\ref{clscal}) follows from the nature of 
the asymptotic future in our $\Lambda$CDM model.  Observers in a universe 
with positive $\Lambda$ have a future event horizon, \ie\ the conformal 
time converges to a finite constant $\eta_f$ as proper time $t \ra \infty$.  
Therefore the angular scaling relation (\ref{lscal}) tells 
us that as proper time (or scale factor) approaches infinity, the $\ell$ 
value for any particular feature in the $C_\ell$ spectrum, such as a peak 
position, will approach a finite maximum, \ie\ features will approach a 
non-zero minimum angular size.  (Geometrically, the LSS sphere approaches a 
maximum comoving radius, so features on it must approach a minimum size.)  
For our fiducial model we chose $\Omega_\Lambda = 0.77$, and so 
\eq(\ref{lscal}) gives for the limiting scaling relation
\beq
\ell_f = \ell_0 \fr{\int_{a_{\rm LS}}^\infty (\ad a)^{-1}da}
                   {\int_{a_{\rm LS}}^{a_0}  (\ad a)^{-1}da} = 1.31\ell_0.
\label{ellfinal}
\eeq
For example, the first acoustic peak, which we observe to be at the 
position $\ell_0 = 221$ today, will asymptote to $\ell_f = 290$ in the 
late de~Sitter phase.  This asymptotic behaviour is in marked contrast to 
that of a purely matter-dominated Einstein-de~Sitter model.  In the 
vanishing $\Lambda$ case, the numerator in \eq(\ref{ellfinal}) diverges 
(no event horizon exists) and the structure in the $C_\ell$ spectrum 
shifts to ever smaller scales.

   The geometrical scaling relation (\ref{clscal}) is very closely related 
to the well-known geometrical parameter degeneracy in the CMB anisotropy 
spectrum between spatial curvature and $\Lambda$ \cite{bet97,zss97,eb99}:  If 
two cosmological models share the same primordial power spectrum $\PR(k)$, 
the same physical baryon and CDM densities today, $\rho_b$ and 
$\rho_c$, and finally the same angular diameter distance $d_A$ 
to the LSS, then they will exhibit essentially identical primary $C_\ell$ 
spectra.  The degeneracy can only be broken by secondary sources of 
anisotropy, such as the integrated Sachs-Wolfe effect, or by other 
cosmological observations.

   To understand the origin of this degeneracy and its relation to the 
preceeding discussion, recall that the energy density in the CMB today, 
$\rho_\gamma$, is fixed to very high accuracy by the measurement of the 
mean temperature, as we mentioned in the Introduction.  Therefore if we 
consider models with identical values of the densities $\rho_b$ and 
$\rho_c$ today, then the densities of baryons, CDM, and 
photons at last scattering are the same for all such models, since the 
densities scale in a well-defined manner (for example, 
$\rho_\gamma/\rho_c \propto a_0/a$).  Therefore, given the same 
initial conditions in the form of $\PR(k)$, models that have common values 
of $\rho_b$ and $\rho_c$ today will have identical local physics 
at least to the time of recombination, when any spatial curvature or 
$\Lambda$ will have negligible effect.  Thus these models will produce 
identical primary anisotropies.

   If the models have different values of $\Omega_K$ (spatial curvature) 
and $\Lambda$, 
then the dynamics, including the propagation of CMB anisotropies, will 
differ significantly at late times as those components come to dominate.  
However, if the models share the same angular diameter distance, then 
their $C_\ell$ spectra, which should be calculated using \eq(\ref{cldef}) 
with $\rls$ replaced by $d_A$ (at least for small scales where the 
effects of spatial curvature on the primordial spectra can be ignored), will 
be identical.  Geometrically, models with identical $\PR(k)$, $\rho_b$, 
and $\rho_c$ share the same local physics to recombination, and hence 
the same physical scales for acoustic wave structures (in particular the 
same sound horizon).  For models which additionally have identical 
$d_A$, 
observers see the anisotropies generated on a spherical shell at the time 
of last scattering of {\em identical\/} physical surface area (given by 
$4\pi d_A^2$).  Hence those physical acoustic scales are mapped to 
identical angular scales in the sky for the different models.  In short, 
the observed primary anisotropies in models with identical $\PR(k)$, 
$\rho_b$, $\rho_c$, and $d_A$ are produced under 
the same local physical conditions on a sphere of identical physical size, 
and hence appear identical.  The scaling relation (\ref{clscal}) describes 
how the observed anisotropies change if we hold the local physics at 
recombination (together with $\Lambda$ and $\Omega_K$) constant, but allow the 
time of observation to vary, which amounts to simply varying the size of the 
sphere at last scattering that generates the observed anisotropies.  The 
parameter degeneracy states that the same anisotropy spectrum can be 
produced even if $\Lambda$ and $\Omega_K$ vary, as long as the size of 
the last scattering sphere is held constant.

   To close this discussion of the primary anisotropies, we introduce 
the power spectrum difference $\delta C_\ell(\eta) \equiv 
C_\ell(\eta') - C_\ell(\eta)$ between the spectra observed at two 
different times.  This is a measurable quantity which we might consider 
a candidate for detecting the evolution of the CMB.  
Given some spectrum $C_\ell(\eta)$ at a single time $\eta$ we can readily 
calculate the difference $\delta C_\ell(\eta)$ using the scaling 
relation (\ref{clscal}).  For small $\delta\eta \equiv \eta' - \eta$, we have
\beq
\delta C_\ell(\eta) \simeq \fr{\pd}{\pd\eta}C_\ell(\eta) \delta\eta,
\label{deltacl}
\eeq
so that the change in the CMB power spectrum at fixed $\ell$ is proportional 
to $\delta\eta$.  As we will see in the next Section, this behaviour differs 
from that of the power spectrum of the difference $\alm(\eta') - \alm(\eta)$.  
Using \eq(\ref{clscal}) for the time dependence, we can write
\beq
\delta\Cl(\eta) = -\fr{\del\eta}{\eta_{\rm LS}} 
   \ld[\ell\fr{\pd\Cl(\eta)}{\pd\ell} + 2C_\ell(\eta)\rd],
\label{deltaclexpl}
\eeq
at first order in $\del\eta/\eta_{\rm LS}$.  Note that $\del\Cl$ can have 
either sign, and will equal zero whenever $\pd(\ell^2\Cl)/\pd\ell = 0$, as 
for example on a scale-invariant Sachs-Wolfe plateau or at an acoustic peak.

   Recall that the quantity $\Cl(\eta)$ describes the relative anisotropies 
$\del T/T$, and hence is insensitive to the bulk expansion redshift.  If 
we wish to consider instead the evolution of the absolute temperature 
anisotropies $\del T$, the relevant quantity to calculate is
\beq
\delta[T^2\Cl(\eta)] = T^2\ld[-2\fr{\del\eta}{(aH)^{-1}}\Cl(\eta)
   + \delta\Cl(\eta)\rd]
\label{deltat2cl}
\eeq
at lowest order in $\del\eta/(aH)^{-1}$, 
where we have used the relation $\dot T = -HT$.  Since in standard 
$\Lambda$CDM models $\eta_{\rm LS}$ is a few times the comoving 
Hubble radius $(aH)^{-1}$, the first term on the left-hand side of 
\eq(\ref{deltat2cl}) 
is of the same order as the second term.  That is, the expansion cooling 
effect is on the same order as the geometrical scaling effect, and so it 
will be important to distinguish the two processes.


\subsection{Integrated Sachs-Wolfe effect} \label{sec:isw}

   The simple scaling relation derived in the previous subsection 
determines the time evolution of the power spectrum of anisotropies 
produced near the LSS.  However, in the standard $\Lambda$CDM model, 
significant anisotropies are also produced at late times as a result of 
the changing equation of state as the Universe becomes cosmological constant 
dominated.  This process is known as the (late) integrated Sachs-Wolfe 
(ISW) effect \cite{sw67,ks85}.  Since these anisotropies are 
produced relatively locally, their time dependence must be explicitly 
calculated.  Note that anisotropies are also generated by the {\em early\/} 
ISW effect, during the time that radiation still significantly contributes 
to the dynamics.  However, those anisotropies are produced relatively 
close to the LSS, adding coherently to the primary Sachs-Wolfe 
contribution, and hence scale as do the primary anisotropies, according 
to \eq(\ref{clscal}).

   The contribution of the late ISW effect can be described by adding to the 
transfer function, \eq(\ref{tdef}), a term $T_{\rm ISW}(k,\ell,\eta)$ which is 
an integral over the line of sight to the LSS.  For the case of interest, 
for which the anisotropic stress is negligible, we have \cite{dodelson03}
\beq
T_{\rm ISW}(k,\ell,\eta) = 2A_3(k)\int_{\eta_{\rm LS}}^\eta d\eta' g'(\eta') 
                           j_\ell[k(\eta - \eta')].
\label{iswint}
\eeq
Here $g'(\eta) \equiv dg/d\eta$, with $g(\eta)$ being the {\em growth 
function\/} 
which describes the temporal evolution of the zero-shear or longitudinal 
gauge curvature perturbation, $\psi$ (also called the ``Newtonian 
potential''), via
\beq
\psi(k,\eta) \equiv g(\eta)A_3(k)\R(k).
\label{gdef}
\eeq
The function $A_3(k)$ is defined such that $g(\eta) \ra 1$ at early times 
in matter domination.  
Then, a gauge transformation between the comoving, $\R$, and zero-shear, 
$\psi$, curvature perturbations during the matter dominated period gives 
$A_3(k) = -3/5$.

   To evaluate the ISW contribution, we need to first determine the evolution 
of the curvature perturbation $\psi$.  To do this, we only need to solve 
the space-space, or dynamical, linearized Einstein equation.  For the case 
where pressure and anisotropic stress perturbations can be ignored, which 
holds in a universe containing only dust and $\Lambda$, this 
equation becomes
\beq
\ddot\psi + 4H\dot\psi + (3H^2 + 2\dot H)\psi = 0
\eeq
(see, \eg, Ref. \cite{mfb92}).  There are no spatial gradients in this 
equation, which confirms that the 
growth function is independent of $k$.  It is straightforward to verify 
that the growing mode solution to this equation is
\beq
\psi(\eta) \propto \fr{H}{a} \int^\eta d\eta' \fr{a^2\dot H}{H^2}.
\eeq
Next, using the relation $a^3\dot H = \const$, which holds exactly for a 
universe consisting of dust and cosmological constant, employing 
\eq(\ref{gdef}), and matching the growing mode solution to the initial 
condition $\psi_0(k) = -(3/5)\R(k)$, we obtain
\beq
g(\eta) = \fr{5}{2}\fr{\Omega_0H}{a} \int^\eta \fr{d\eta'}{aH^2},
\label{growth}
\eeq
where $\Omega_0$ is the density in matter today relative to the critical 
density, and the scale factor is normalized to unity today.

   Now that the evolution of the growth function has been determined, 
we can evaluate the ISW contribution to the power spectrum.  It can be 
shown that the cross term between the Sachs-Wolfe and (late) ISW terms is 
negligible, so the two add incoherently in $\Cl$.  For the ISW part we have
\bea
C_\ell^{\rm ISW}(\eta) &=& 4\pi\int\fr{dk}{k}\PR(k)T_{\rm ISW}^2(k,\ell,\eta)\\
   &=& \fr{72}{25}\fr{\pi^2\PR}{\ell^3}
       \int_0^\eta d\eta' g'^2(\eta')(\eta - \eta').
\label{isw}
\eea
To obtain this result we have assumed a scale invariant primordial spectrum, 
$\PR(k) = \PR$, and we have used the relation \cite{ks94}
\beq
\int_0^\infty \fr{dk}{k} j_\ell[k(\eta - \eta')] j_\ell[k(\eta - \eta'')]
\simeq \fr{\pi}{2\ell^3} (\eta - \eta')\delta(\eta' - \eta''),
\label{iswapprox}
\eeq
which holds for large $\ell$.  Unfortunately it is at small $\ell$ that the 
ISW effect is greatest, so this approximation, which appears to be the best 
we can do analytically, is not terribly accurate at very late times, when, 
as we shall see, the ISW contribution becomes very large.  Nevertheless, 
\eq(\ref{isw}) will give a reasonable estimate of the ISW contribution to 
a change $\del\Cl$ over short time intervals.

   Given a primordial amplitude, $\PR$, and a matter density parameter today, 
$\Omega_0$, \eqs(\ref{growth}) and (\ref{isw}) allow us to calculate the 
ISW contribution to the power spectrum at any time.  In addition, taking 
the time derivative of \eq(\ref{isw}), we find for the rate of change of 
the ISW contribution
\beq
\fr{\pd}{\pd\eta}C_\ell^{\rm ISW}(\eta) = \fr{72}{25}\fr{\pi^2\PR}{\ell^3}
       \int_0^\eta d\eta' g'^2(\eta').
\eeq
Therefore, combining this expression with \eq(\ref{deltaclexpl}) for the 
change in the primary Sachs-Wolfe power spectrum over short time intervals 
$\del\eta$, we find for the total contribution
\bea
\delta\Cl(\eta) = 
   &-&\fr{\del\eta}{\eta_{\rm LS}}
      \ld[\ell\fr{\pd\Cl(\eta)}{\pd\ell} + 2\Cl(\eta)\rd.\nonumber\\
   &-& \ld.\fr{72}{25}\fr{\pi^2\PR}{\ell^3}\eta_{\rm LS}
           \int_0^\eta d\eta' g'^2(\eta')\rd].
\label{deltaclisw}
\eea


\subsection{Gravitational waves}
\label{sec:tensors}

   Inflationary models generically predict a spectrum of primordial 
gravitational waves, although the relative contribution of these tensor modes 
to the total CMB anisotropy ranges from substantial to very small, 
depending on the model.   The anisotropies arise through the tensor analogue 
of the scalar ISW line of sight integral, \eq(\ref{iswint}).  In place of 
the time derivative of the scalar curvature perturbation $\psi$, 
the tensor contribution involves the rate of change of the transverse 
and traceless part of the spatial metric perturbation, $h_{ij}$.  The 
evolution of this part of the metric perturbation is given by the 
dynamical Einstein equation, which becomes (see, \eg, \cite{mfb92})
\beq
h_{ij}'' + 2aHh_{ij}' - \nabla^2h_{ij} = 0
\eeq
in the absence of tensor anisotropic stress, which is a valid approximation 
in the matter- or $\Lambda$-dominated regimes.

   The dynamics of $h_{ij}$ 
as dictated by this equation depends on the mode wavelength relative to 
the Hubble scale.  For $k/(aH) \ll 1$, the tensor mode is overdamped, and 
the (growing) mode decays very slowly.  For $k/(aH) \gg 1$, the mode 
undergoes underdamped oscillations, decaying like $h_{ij} \propto 
e^{ik\eta}/a$ in the adiabatic regime.   Therefore, for a particular 
tensor mode $k$, the rate of change $h_{ij}'$ is peaked near the time 
that the mode crosses the Hubble radius, $k/(aH) \sim 1$, and is small at 
early and late times.  This means that the contributions to the CMB 
anisotropies from a particular scale arise primarily when that scale 
enters the Hubble radius.  In particular, scales that enter significantly 
before last scattering will have decayed before they could source the 
CMB.  This imposes a small scale cut-off on the tensor anisotropy spectrum, 
with negligible power for
\beq
\ell \gg \ell_c \equiv \rls\ad_{\rm LS},
\eeq
where $1/\ad_{\rm LS}$ is the comoving Hubble radius at last scattering.  
Since $\ad_{\rm LS}$ is fixed for a particular model, the $\ell$ cut-off 
scales with time of observation according to
\beq
\ell_c(\eta') = \ell_c(\eta)\fr{\rls'}{\rls},
\label{gwcut}
\eeq
which is the same scaling as for primary features in the scalar CMB spectrum, 
\eq(\ref{lscal}).

   For $\ell \ll \ell_c$, detailed calculations for a matter dominated 
universe \cite{star85,turner93} show that the tensor anisotropy spectrum is 
nearly 
flat, mimicing the Sachs-Wolfe plateau.  In this case the tensor spectrum 
does not evolve apart from the scaling of the cut-off, \eq(\ref{gwcut}).  
However, as described above, the largest angular scales will be sourced at 
the latest times, and so for a universe with cosmological constant this will 
lead to some dependence on observation time for those scales as the equation 
of state changes at late times.  For example, for time of observation $\eta$ 
the tensor quadrupole is sourced near very roughly the conformal time 
$\eta/2$ \cite{star85}.  In the matter-dominated era, the comoving Hubble 
radius $1/(aH) = 1/\ad$ increases with time, but as the Universe enters the 
$\Lambda$-dominated era, $1/(aH)$ starts to {\em decrease\/} (indeed this 
defines acceleration).  Therefore, the largest scale modes that contribute 
to the tensor anisotropies will enter the Hubble radius somewhat 
{\em later\/} in the presence of a cosmological constant than without 
(sufficiently large-scale modes will {\em never\/} enter the Hubble radius).  
The modes which enter near time $\eta_f/2$, where $\eta_f$ is the asymptotic 
final conformal time, will have a significantly delayed entry time, and 
therefore we expect the very largest scale tensor anisotropies to be somewhat 
reduced at the latest times.


\subsection{Optical depth}
\label{sec:optdepth}

   One final line of sight effect on the CMB anisotropies that we will 
consider is the time dependent optical depth due to Thomson scattering.  
Looking back, it is the rapid increase in scattering near the time of 
recombination that makes it possible to speak of a ``surface of last 
scattering'', where the primary anisotropies are emitted.  At later 
times, reionization results in a time dependent attenuation of the 
amplitude of the power spectrum, which it will be important to quantify.

   In our fiducial model the epoch of reionization occurs at redshift 
$z_R = 11.1$, or $a_R = 0.083$, with the 
scale factor normalized to unity today.  Here, we see a decrease in the 
amplitude of intermediate to small scale ($\ell \gtrsim 30$) anisotropies 
as photons are rescattered.  On these scales, $C_{\ell}(\eta)$ is reduced 
by a factor $e^{2 \tau_R}$, where $\tau_R$ is the optical depth 
to reionization.  To compute the suppression in $C_{\ell}(\eta)$, recall 
the definition of the optical depth $\tau(a, a_{\rm obs})$ between scale 
factors $a$ and $a_{\rm obs}$, given by
\beq
\tau(a, a_{\rm obs}) = \sigma_{\rm T} \int^{a_{\rm obs}}_{a} 
   \frac{dt}{da^{\prime}}\, n_e (a^\prime) \, da^{\prime}\,,
\label{taudef}
\eeq
where $\sigma_{\rm T}$ is the Thomson scattering cross-section and 
$n_e(a)$ the electron number density.  Assuming reionization is sharp 
and the energy density of radiation is subdominant, the optical depth to 
reionization, $\tau_R(a_{\rm obs}) \equiv \tau(a_R, a_{\rm obs})$ 
is given by (see, \eg, \cite{gbl99})
\bea
\tau_{\rm R}(a_{\rm obs})
 = 0.046(1 - Y_p)\!\!\!\!\!\!\!\!&&\frac{\Omega_b}{\Omega_m}
    \left(\sqrt{\Omega_m h^2 a_R^{-3} + \Omega_{\Lambda}h^2}\rd.\nonumber\\
  &-& \ld.\sqrt{\Omega_m h^2 a_{\rm obs}^{-3} + \Omega_{\Lambda}h^2}\rd),
\label{reiondepth}
\eea
with $\tau_R(a_{\rm obs}) = 0$ for $a_{\rm obs} \leq a_R$.  
In this expression $Y_p$ is the primordial helium fraction (assumed 
to be 0.24).

   \eq(\ref{reiondepth}) gives $\tau_R(1) = 0.088$, so reionization 
reduces $\Cl(\eta)$ by a factor of approximately $0.84$ by today 
[of course we must evaluate the spectrum at different times at the $\ell$ 
values related by the scaling relation \eq(\ref{lscal})].  Much of 
the suppression of $C_{\ell}(\eta)$ occurs before the present time.  
Since the time the scale factor was half its present value, for example, 
the optical 
depth to reionization has only increased by $0.003$.  As $a_{\rm obs} \ra 
\infty$, \eq(\ref{reiondepth}) says that the optical depth will only 
increase by $\del\tau_\infty \equiv \tau_R(\infty) - \tau_R(1) 
= 5.7 \times 10^{-4}$, so that $C_{\ell}(\eta)$ will only be reduced by 
$0.1 \%$ relative to the value today.  (Note that the peak {\em positions} 
should not be effected by reionization, since the peak structure is 
at scales smaller than those corresponding to the particle horizon 
at rescattering.)  Therefore the Universe is essentially transparent on 
cosmological scales today, and we have already seen nearly the maximum 
amount of reionization damping.


\subsection{Time evolution from CAMB}
\label{sec:clcamb}

   In order to confirm and extend the analytic results of previous 
subsections we have 
modified the \camb\ software \cite{lcl00} to compute the CMB power spectrum at 
different observational times. To do this, we simply modify all routines 
such that we can evaluate the transfer functions $T(k,\ell,\rls)$ at 
different $\rls$, using the set of best-fitting cosmological parameters 
as measured today.   The changes required to \camb\ are straightforward---for 
the most part all that is needed is to change the {\textsc{tau0} variable 
to trick the code into using a different observational time.  
The parameters we use are those of a standard six parameter spatially flat 
$\Lambda$CDM model, given by $\Omega_c h^2 = 0.104$, $\Omega_b 
h^{2} = 0.0223$, $h = 0.734$, $n_S = 0.951$, $A_S = 2.02 \times 
10^{-9}$, and $z_R = 11.1$, where $\PR(k) = A_S 
(k/k_0)^{n_S - 1}$ with pivot scale $k_0 = 0.05\,{\rm Mpc}^{-1}$, and 
$z_R$ is the redshift of reionization.  Where necessary, we set 
$n_S = 1$ to simplify comparison with analytic results.  We do not expect 
significant differences in our results if these parameters are varied 
somewhat.  Since the parameters 
are defined relative to their present day values, the spectrum measured 
by an observer with $z > z_R$, for example, will not be affected by 
the epoch of reionization and a future observer will see a universe 
completely dominated by dark energy.

   In Fig.~\ref{fig:cl_a} we plot the power spectrum $\ell(\ell + 1) 
C_{\ell} (\eta)/(2\pi)$ calculated from our modified version of \camb, 
parameterizing the time dependence in terms of the 
observational scale factor $a_{\rm obs}$, where $a_{\rm obs} = 1$ 
corresponds to today.  It is clear that the angular scale of features 
resulting from projection of inhomogeneities near the LSS can be understood 
from the scaling relation~(\ref{clscal}).  For example, in 
Fig.~\ref{fig:cl_a} we show the predicted scaling of the first acoustic 
peak position (located at $\ell = 221$ at the present time) and find 
extremely good agreement with the predicted value.  The existence of a 
future event horizon means that during $\Lambda$ domination $d\rls /
d a_{\rm obs}$ tends to zero and so the acoustic peak positions become 
``frozen in''.  With our parameters we see that the first acoustic peak 
becomes frozen in at the value $\ell_f \simeq 290$ as we predicted in 
\eq(\ref{ellfinal}).

\begin{figure}[ht]
\includegraphics[width=\columnwidth]{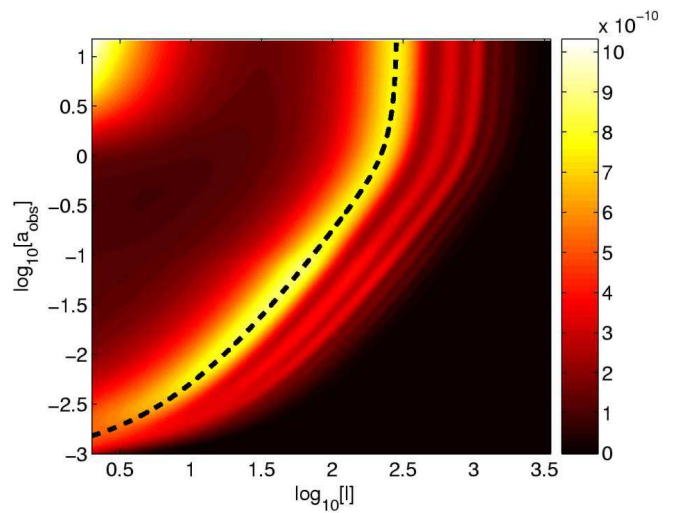}
\caption{\label{fig:cl_a} CMB power spectrum  $\ell(\ell + 1)
C_{\ell}/(2\pi)$ as a function of multipole $\ell$ and scale factor of 
observation $a_{\rm obs}$, as calculated from our modified version of 
\camb\ for our 
fiducial $\Lambda$CDM model.  We also plot the analytic scaling of the 
first acoustic peak (thick dashed line) predicted by \eq(\ref{clscal}).}
\end{figure}

   Recall that the scaling relation \eq(\ref{clscal}) predicts not only how 
the angular sizes of features in the spectrum scale with time, but also 
that the {\em magnitude\/} of the power, $\ell(\ell + 1)C_{\ell}(\eta)$, 
remains constant into the future at, \eg, any acoustic peak.  For late times, 
$a_{\rm obs} \agt 0.3$, \fig\ref{fig:cl_a} indeed confirms this prediction.  
Our fiducial model underwent reionization at $a_R = 0.083$, and we 
predicted in Section~\ref{sec:optdepth} that as a result the power 
spectrum should be attenuated by approximately $16\%$ on all but the 
largest scales.  Again, this is visible in \fig\ref{fig:cl_a}.  Recall 
that we predicted a negligible $0.1\%$ reduction in $\Cl(\eta)$ between 
today and the distant future.

   Also visible in \fig\ref{fig:cl_a} is a substantial increase in power at 
the largest scales at late times due to the increasing ISW effect in 
our $\Lambda$CDM model, which we discussed in Section~\ref{sec:isw}.  
Indeed, for $a_{\rm obs} \gtrsim 5.0$ the quadrupole power actually 
exceeds the power at the first acoustic peak.  The ISW contribution converges 
as the observation scale factor $a_{\rm obs}$ approaches infinity, since 
in the integral in \eq(\ref{isw}), we have $g'(\eta) \ra 0$ and $\eta \ra 
\eta_f$ as $a_{\rm obs} \ra \infty$, where $\eta_f$ is finite.  The 
asymptotic form of the power spectrum at late times is plotted in 
\fig\ref{fig:futscalten}, together with the current spectrum.  The dramatic 
increase in the ISW contribution, as well as the shift in peak positions 
predicted in \eq(\ref{ellfinal}), are clearly visible.  
Note that the ISW contribution to the dipole power, though not shown in 
\fig\ref{fig:futscalten}, asymptotes to $C_1^{\rm ISW}(\eta_f) 
\simeq 5\times10^{-9}$ for out model, according to \camb.  This corresponds 
to typical dipole ISW amplitudes $a_{1m}^{\rm ISW} \simeq \sqrt{C_1^{\rm ISW}} 
\simeq 7\times10^{-5}$.  Although such amplitudes represent a large increase 
over the current ISW dipole, they are still considerably below either the 
current total measured dipole, $a_{10} = 2.5\times10^{-3}$ (with the polar 
axis aligned with the dipole), or the amplitude of the galactic orbital 
dipole oscillation, $a_{10} = 1.5\times10^{-3}$ according to 
\eq(\ref{eqn:suncmb}).

\begin{figure}[ht]
\includegraphics[width=\columnwidth]{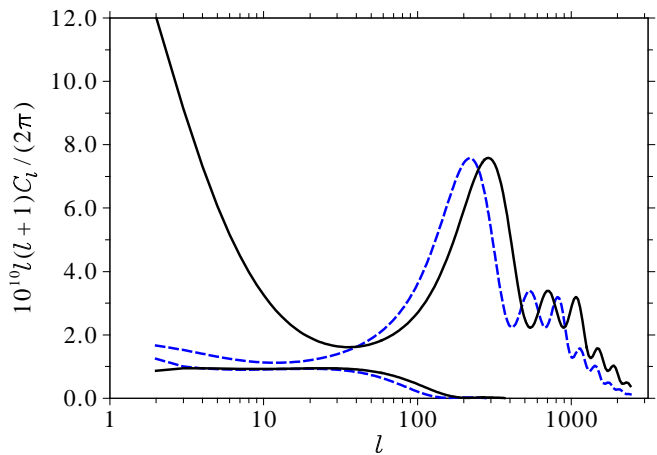}
\caption{Anisotropy power spectra for scalars (top pair of curves) and 
for tensors (bottom pair, arbitrary scale), calculated using our modified 
version of \camb\ for today (dashed curves) and for the asymptotic future 
(solid).}
\label{fig:futscalten}
\end{figure}

   \fig\ref{fig:futscalten} also presents the gravitational wave contribution 
to the anisotropy spectra today and in the asymptotic future.  In 
Section~\ref{sec:tensors} we described the expected behaviour of tensor 
modes in the future, which entailed the same geometrical scaling of the 
small-scale cut-off in the spectrum, as well as a decrease in power at 
the very largest scales.  Both of these features are visible in 
\fig\ref{fig:futscalten}.

   As a check on our custom modifications to \camb, we plot in 
\fig\ref{fig:pscambanal} the power spectrum from \camb\ for 
$a_{\rm obs} = 2.0$, as well as the corresponding curve calculated from 
a power spectrum generated for today, $a_{\rm obs} = 1$, and transformed 
to $a_{\rm obs} = 2.0$ using the scaling relation \eq(\ref{clscal}).  
Additionally, the spectrum calculated from the scaling relation includes 
the increased ISW component calculated from the analytical approximation 
\eq(\ref{isw}).  To facilitate the use of this analytical expression, the 
spectral index was set to $n_S = 1$ for these calculations.  Since the 
curves coincide at all but the largest scales, 
it is clear that the scaling relation has accurately captured the evolution 
of $\Cl(\eta)$.  However, the ISW contribution is substantially 
overestimated, indicating the limitations of the approximation 
\eq(\ref{iswapprox}) involved in deriving \eq(\ref{isw}).

\begin{figure}[ht]
\includegraphics[width=\columnwidth]{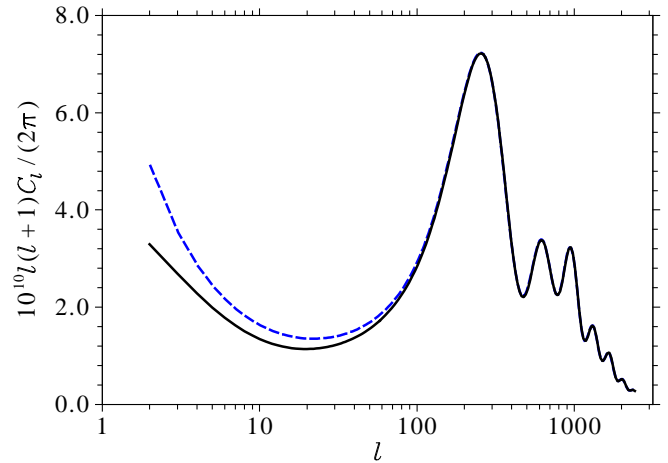}
\caption{Anisotropy power spectrum for $a_{\rm obs} = 2.0$ calculated with 
our modified version of \camb\ (solid line) and using the scaling relation 
\eq(\ref{clscal}) and analytical ISW approximation \eq(\ref{isw}) 
(dashed).  We set $n_S = 1$ for these calculations.}
\label{fig:pscambanal}
\end{figure}

   To make further contact with the analytic results in previous subsections, 
in Fig.~\ref{fig:cldiff} we plot the difference $\del\Cl \equiv 
\Cl(a'_{\rm obs}) - \Cl(a_{\rm obs})$ calculated using our modified 
version of \camb\ between 
the power spectrum today, at $a_{\rm obs} = 1$, and at a future time, when 
$a'_{\rm obs} = 1 + \delta a$, for the cases $\delta a = 10^{-4}$, $0.001$, 
and $0.01$.  These curves exhibit very accurately the scaling with 
$\delta a$ predicted in \eq(\ref{deltacl}), when we recall that $\delta a 
= H_0\delta\eta$ for small $\delta\eta$.  Slight departures from this 
simple scaling are evident at the largest scales, where the spectra are 
nearly flat and hence their precise shape sensitively influences the location 
of zeros in $\del\Cl$.  In \fig\ref{fig:cldiff} we also plot the 
analytical result calculated from the power spectrum today using 
\eq(\ref{deltaclisw}).  Again we find excellent agreement with \camb\ at 
all but the largest scales.  We also find reasonable agreement at low 
$\ell$, indicating that our ISW approximation is quite good for small 
time increments from today.

\begin{figure}[ht]
\includegraphics[width=\columnwidth]{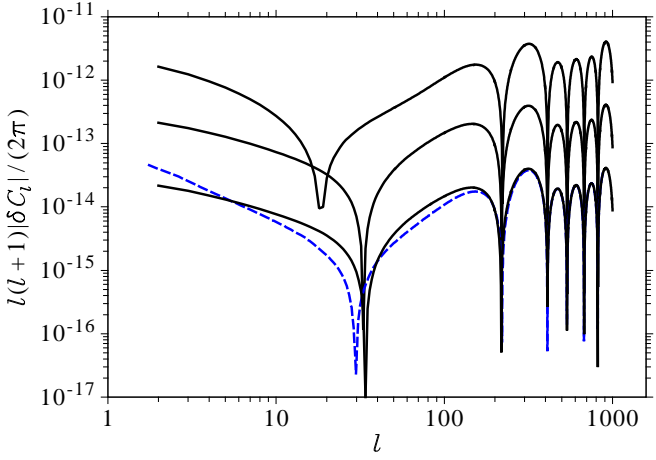}
\caption{Absolute value of the difference in the CMB power spectra 
$\ell(\ell + 1)|\delta\Cl|/(2\pi)$ between $a_{\rm obs} = 1$ and $a'_{\rm obs}
 = 1 + \delta a$.  The solid curves were calculated from our modified version 
of \camb, and from top to bottom denote $\delta a = 0.01$, $0.001$, and 
$10^{-4}$.  The dashed curve was calculated using the analytical 
expression \eq(\ref{deltaclisw}) for $\delta a = 10^{-4}$.  We used 
$n_S = 1$ for these calculations.}
\label{fig:cldiff}
\end{figure}


\section{The difference map power spectrum}
\label{sec:diffmap}

\subsection{Analytical time evolution}
\label{sec:analdiffmap}

   In the previous section we found a very simple scaling relation to describe 
the time dependence of the CMB power spectrum, which, together with the ISW 
effect, thoroughly describes the evolution of the spectrum.  However, if we 
are interested in the best way to observe evolution in the CMB we might expect 
that observing changes in the actual sky map, or the $\alm$s, should 
be far more promising than looking for changes in the heavily compressed 
$\Cl$ power spectrum.  Intuitively, as the shell $r = \rls$ of the LSS 
grows in size, we expect the finest structures to change first, then 
the larger ones.  As we shall see, the difference between two sky maps 
measured at different times does indeed encode much more information 
than the $\Cl$ spectra, namely the {\em correlations\/} between the two 
maps, although perhaps counterintuitively the {\em magnitude\/} of 
a change $\Cl$ will dominate over the difference map power spectrum for 
small time intervals.

\subsubsection{Definitions}

   Consider two measurements of the $\alm$s at times $\eta$ and 
$\eta'$ and define the difference map by
\beq
\delta\alm \equiv \alm(\eta') - \alm(\eta).
\label{diffmapdef}
\eeq
Using \eq(\ref{almrtr}), we can readily calculate the statistical 
properties of the difference map.  We find
\beq
\bra\delta\alm\delta a^\ast_{\ell'm'}\ket
   = D_\ell^{\eta\eta'}\delta_{\ell\ell'}\delta_{mm'},
\label{diffmapps}
\eeq
where we define the power spectrum of the difference map, 
$D_\ell^{\eta\eta'}$, by
\beq
D_\ell^{\eta\eta'} \equiv \Cl(\eta) + \Cl(\eta') - 2\Cl^{\eta\eta'},
\label{dmpsdef}
\eeq
and
\beq
\Cl^{\eta\eta'} \equiv 4\pi\int\fr{dk}{k}\PR(k)T(k,\ell,\rls)T(k,\ell,\rls').
\label{corfnc}
\eeq
Note that the quantity $\bra\delta\alm\delta a^\ast_{\ell'm'}\ket$ is 
diagonal in $\ell$ and $m$, and that $\Cl^{\eta\eta} = \Cl(\eta)$.

   The quantity $\Cl^{\eta\eta'}$ is an unequal time correlator, \ie\ a 
correlation function that relates 
the anisotropies at time $\eta$ with those at time $\eta'$, through
\beq
{\rm Re}\,\bra\alm(\eta)a^\ast_{\ell'm'}(\eta')\ket
   = \Cl^{\eta\eta'}\delta_{\ell\ell'}\delta_{mm'}.
\label{corfncdef}
\eeq
Since the variances $\Cl(\eta)$ and $\Cl(\eta')$ will in general differ, 
the quantity $\Cl^{\eta\eta'}$ is not the best measure of correlations, 
and the spectrum $D_\ell^{\eta\eta'}$ measures not only the loss of 
correlations but also the change in variance $\Cl(\eta)$.  Therefore we 
may consider instead the modified difference map
\beq
\overline{\delta a}_{\ell m}
   \equiv \sqrt{\fr{\Cl(\eta)}{\Cl(\eta')}}\alm(\eta') - \alm(\eta),
\label{diffmapdefalt}
\eeq
which normalizes the modes at $\eta'$ to have the same variance as those 
at $\eta$.  Then we find
\beq
\bra\overline{\delta a}_{\ell m}\overline{\delta a}{}^\ast_{\ell'm'}\ket = 
 2\Cl(\eta)\ld(1 - \bar{C}_\ell^{\eta\eta'}\rd)\delta_{\ell\ell'}\delta_{mm'},
\label{diffmappsalt}
\eeq
where we have defined the normalized correlation function by
\beq
\bar C_\ell^{\eta\eta'} \equiv
   \fr{C_\ell^{\eta\eta'}}{\sqrt{\Cl(\eta)\Cl(\eta')}}.
\label{normcorfnc}
\eeq
This normalized function is useful in that we have $\bar C_\ell^{\eta\eta'} 
= 1$, $0$, and $-1$ for perfect correlations, no correlations, and perfect 
anticorrelations, respectively.  Similarly, the quantity $1 - 
\bar C_\ell^{\eta\eta'}$ measures the loss of correlations alone.  However, 
the spectrum $D_\ell^{\eta\eta'}$ will still be useful, since it measures 
both the loss of correlations and the change in variance, so it might be 
expected to be more sensitive to changes in the CMB than the quantity 
$1 - \bar C_\ell^{\eta\eta'}$.  
Also, through the definition (\ref{diffmapdef}), the quantity 
$D_\ell^{\eta\eta'}$ is more directly tied to observations.

\subsubsection{Time evolution---flat sky approximation}

   In analogy with \eq(\ref{deltacl}) for $\delta\Cl$, we can 
write the spectrum of the difference map for small increments in time 
$\del\eta = \eta' - \eta$ as
\beq
D_\ell^{\eta\eta'} \simeq
 \ld\bra\fr{\pd\alm}{\pd\eta}\fr{\pd\alm^\ast}{\pd\eta}\rd\ket (\delta\eta)^2.
\label{difftdep}
\eeq
Therefore the difference of the power spectrum, $\delta\Cl$, dominates 
over the power spectrum of the difference map, $D_\ell^{\eta\eta'}$, for 
small enough $\delta\eta$, since $\delta\Cl$ is only proportional to the 
first power of $\delta\eta$.  Of course for a particular $\delta\eta$ 
we must calculate the coefficients of $\delta\eta$ and $(\delta\eta)^2$ 
before we decide which method is more efficient if we are interested in 
a detection.  The details of instrumental noise are important and this 
is discussed fully in \cite{mzs07}.

   Beyond the $(\delta\eta)^2$ scaling, it is much more difficult to obtain 
the detailed evolution of $D_\ell^{\eta\eta'}$ than it was 
for $\delta\Cl$.  Even when we consider only the Sachs-Wolfe plateau 
contribution, for which $A_1(k) = -1/5$ and $A_2(k) = A_3(k) = 0$, the 
Bessel integrals involved in \eq(\ref{corfnc}) cannot be analytically 
solved.  In fact, this problem is related to a divergence that can be 
illuminated if we employ the flat sky approximation described in the 
\hyperref[flatskyapp]{Appendix}.

   Under that approximation, which is valid over small patches of sky and 
replaces the discrete indices $\ell$ and $m$ with the continuous 
two-dimensional vector $\bl$, and the polar coordinate $r$ with a Cartesian 
coordinate $x$ parallel with the line of sight, we can readily calculate 
the quantity on the left-hand side of \eq(\ref{diffmapps}).  Using
\beq
\delta a(\bl) \equiv a(\bl,\xls') - a(\bl,\xls)
\eeq
to define the difference map, where $\xls$ and $\xls'$ are the comoving 
distances to the LSS at times $\eta$ and $\eta'$, respectively, and 
using \eq(\ref{almrtrfs}) for the anisotropies, 
we find that the result is not diagonal in $\bl$.  Rather, it contains 
terms proportional to the Dirac functions $\delta^2(\alpha\bl - \bl\,')$ and 
$\delta^2(\alpha^{-1}\bl - \bl\,')$, where we have defined
\beq
\alpha \equiv \fr{\xls'}{\xls}.
\eeq
Indeed this is not surprizing: in the 
flat sky approximation, an anisotropy on angular scale $\bl$ at $\xls$ 
corresponds to a physical mode with comoving wavevector component 
$\bl/\xls$ orthogonal to the line of sight.  But \eq(\ref{primps}) tells 
us that such a mode should share correlations with the {\em same\/} 
physical scale at $\xls'$, which corresponds to the angular scale 
$\bl\xls'/\xls$.  Such off-diagonal correlations are completely suppressed 
in the full spherical expansion, as we found.

   The relevant quantity to calculate in the flat sky approximation is 
instead the power in the difference map defined by
\beq
\wtl{\delta a}(\bl) \equiv a(\alpha\bl,\xls') - a(\bl,\xls).
\label{diffmapdeffs}
\eeq
Again applying \eq(\ref{almrtrfs}) for $a(\bl,\xls)$, we find
\beq
\ld\bra\wtl{\delta a}(\bl)\wtl{\delta a}{}^\ast(\bl\,')\rd\ket
   = D^{\eta\eta'}\!(\bl)\,\delta^2(\bl - \bl\,'),
\eeq
which {\em is\/} diagonal in $\bl$, with the power spectrum of the 
difference map given by
\beq
D^{\eta\eta'}\!(\bl) \equiv C(\bl,\eta) + \alpha^{-2}C(\alpha\bl,\eta')
   - 2C^{\eta\eta'}\!(\bl).
\label{dmpsdeffs}
\eeq 
Here $C(\bl,\eta)$ is the flat sky approximation to the anisotropy power 
spectrum, given by \eq(\ref{cldeffs}), and $C^{\eta\eta'}\!(\bl)$ is the 
correlation function given by
\beq
C^{\eta\eta'}\!(\bl) \equiv \fr{\pi}{\xls'^2}\!\int_{-\infty}^\infty\!\!\!\!\!
   dk_x\!\fr{\PR(k)|T(k,k_x)|^2\cos(k_x\del\xls)}{k^3},
\label{corfncfs}
\eeq
where $T(k,k_x)$ is the flat sky transfer function, $\del\xls \equiv 
\xls' - \xls$, and $k_x$ is the component of the comoving wavevector 
parallel to the line of sight.  \eqs(\ref{diffmapdeffs}) to (\ref{corfncfs}) 
are the flat sky analogues of \eq(\ref{diffmapdef}) to (\ref{corfnc}), 
respectively.  (The continuous argument $(\bl)$ will always distinguish 
quantities in the flat sky approximation from the corresponding exact 
quantities, which are labelled with the discrete indices ${}_{\ell m}$.)

   Note that the integrand in \eq(\ref{corfncfs}), which is exact apart from 
the flat sky approximation, is bounded in magnitude by the integrand in 
\eq(\ref{cldeffs}) for the power spectrum $C(\bl,\eta)$, as we vary 
$\del\xls$.  In place of \eq(\ref{normcorfnc}), the normalized correlation 
function becomes in the flat sky approximation
\beq
\bar C^{\eta\eta'}\!(\bl) \equiv
   \fr{C^{\eta\eta'}\!(\bl)}{\sqrt{C(\bl,\eta)\alpha^{-2} C(\alpha\bl,\eta')}},
\eeq
which is bounded by $|\bar C^{\eta\eta'}\!(\bl)| \le 1$.  In the limit of 
short time interval, $\del\xls \ra 0$, we have $\bar C^{\eta\eta'}\!(\bl) 
\ra 1$, corresponding to perfect correlation.  We also have 
$\bar C^{\eta\eta'}\!(\bl) \ra 0$ as $\del\xls \ra \infty$ (recall, however, 
that in a $\Lambda$CDM universe, only finite conformal time is available 
into the future).

\subsubsection{Special cases}

   Armed with the above flat sky approximation, we can now calculate the 
difference map power spectrum and correlation function in some special 
cases.  First, consider the short time interval case, $\del\xls \ra 0$.  
Expanding \eq(\ref{dmpsdeffs}) in powers of $k_x\del\xls \sim 
\ell\del\xls/\xls$, where $\ell \equiv |\bl|$, we find
\beq
D^{\eta\eta'}\!(\bl) = \pi\ld(\fr{\delta\xls}{\xls}\rd)^2\!\!
   \int_{-\infty}^\infty\!\! dk_x \fr{\PR(k)|T(k,k_x)|^2k_x^2}{k^3}
\label{dmpsfssmallt}
\eeq
for the power spectrum of the difference map at lowest order in 
$\ell\del\xls/\xls$.  This expression exhibits 
precisely the time interval dependence that we predicted in 
\eq(\ref{difftdep}).  (Note that in defining the difference map through 
\eq(\ref{diffmapdeffs}), we have fixed the observed transverse wavevectors 
at both observation times, so the integrand in \eq(\ref{dmpsfssmallt}) 
is independent of $\del\eta$.)

   The integrand in \eq(\ref{dmpsfssmallt}) resembles closely that for 
the anisotropy power spectrum in \eq(\ref{cldeffs}), but with an extra 
factor of $k_x^2$ in the numerator.  In fact, using the relation
\beq
k^2 = k_x^2 + \ld(\fr{\ell}{\xls}\rd)^2
\eeq
we can easily rewrite \eq(\ref{dmpsfssmallt}) as
\beq
D^{\eta\eta'}\!(\bl) = \ld(\fr{\delta\xls}{\xls}\rd)^2\!
   \ld[\ell_0^2 C^{(n_S + 2)}(\bl,\eta) - \ell^2 C(\bl,\eta)\rd].
\label{dmpsfssmallt2}
\eeq
Here $C^{(n_S + 2)}(\bl,\eta)$ is the anisotropy power spectrum calculated 
using a modified primordial power spectrum  defined by
\beq
\p_{\cal R}^{(n_S + 2)}(k) \equiv \ld(\fr{k}{k_0}\rd)^2\PR(k),
\eeq
where $k_0 \equiv \ell_0/\xls$ is the ``pivot scale'' used to define the 
primordial spectrum.  (The result for $D^{\eta\eta'}\!(\bl)$ is, of course, 
independent of the 
pivot scale chosen.)  For the special case of a power law primordial 
spectrum $\PR(k)$, with scalar spectral index $n_S$, the modified spectrum 
$\p_{\cal R}^{(n_S + 2)}(k)$ has spectral index $n_S + 2$; hence our choice 
of notation.  \eq(\ref{dmpsfssmallt2}) says that, for small time increments, 
the {\em shape\/} of the power spectrum of the difference map is 
determined entirely by the actual anisotropy spectrum ``blue tilted'', \ie\ 
$\ell^2C(\bl,\eta)$, together with the spectrum $C^{(n_S + 2)}(\bl,\eta)$ 
calculated 
from a blue-tilted primordial spectrum, both evaluated at the {\em same 
time\/} $\eta$.  Therefore we expect that generically the shape of the 
difference map spectrum $D^{\eta\eta'}\!(\bl)$ will be roughly that of a 
strongly blue-tilted version of the anisotropy spectrum $C(\bl,\eta)$.  The 
{\em height\/} of the spectrum of the difference map is determined by the 
ratio $\del\xls/\xls$.

   Next, we can specialize to the case of the pure scale-invariant ($n_S = 1$) 
Sachs-Wolfe plateau, which is characterized by $T(k,k_x) = A_1(k) = \const$ 
and $\PR(k) = \const$.  \eq(\ref{cldeffs}) gives in this case
\beq
C(\bl,\eta) = \fr{2\pi\PR A_1^2}{\ell^2},
\label{swfs}
\eeq
in agreement with the standard Sachs-Wolfe result, to order $1/\ell$.  The 
normalized correlation function is then
\beq
\bar C^{\eta\eta'}\!(\bl) = \fr{\ell^2}{2\xls^2} 
                    \int_{-\infty}^\infty dk_x \fr{\cos(k_x\del\xls)}{k^3}.
\label{corfncfssw}
\eeq

   In the short time interval limit, $\ell\delta\xls/\xls \ll 1$, 
\eq(\ref{dmpsfssmallt}) becomes for the Sachs-Wolfe plateau
\beq
D^{\eta\eta'}\!(\bl) = \pi\PR A_1^2\ld(\fr{\delta\xls}{\xls}\rd)^2
                     \int_{-\infty}^\infty dk_x \fr{k_x^2}{k^3}.
\label{corfncshortt}
\eeq
Note that this last integral is logarithmically divergent, but this is 
just an artifact of our assumption of a scale invariant spectrum 
to arbitrarily small scales \footnote{The integral in the exact expression, 
\eq(\ref{corfncfssw}), is not divergent, so more fundamentally the 
divergence in \eq(\ref{corfncshortt}) is due to our truncation of the 
series expansion for the cosine in (\ref{corfncfssw}).}.  Equivalently, 
\eq(\ref{dmpsfssmallt2}) cannot be applied in this case, because the 
Sachs-Wolfe integral diverges for $n_S \ge 3$.  In reality, damping within 
the LSS imposes an effective cut-off, with essentially no structure at 
wavenumbers above some value $k_{\rm max}$ \footnote{Of course the 
Sachs-Wolfe plateau for an {\em actual\/} spectrum $D^{\eta\eta'}(\bl)$ will 
receive contributions from the full acoustic peak structure, so the details 
of the cut-off procedure are irrelevant here.}.  Replacing the infinite 
limits with $\pm k_{\rm max}$, we can evaluate the integral in 
\eq(\ref{corfncshortt}) with the result (valid for $\ell/\xls \ll 
k_{\rm max}$)
\beq
D^{\eta\eta'}\!(\bl) \simeq 2\pi\PR A_1^2\ld(\fr{\delta\xls}{\xls}\rd)^2
                            \!\ld(\ln\fr{2k_{\rm max}\xls}{\ell} - 1\rd).
\label{corfncsw}
\eeq
This means that the contribution to the difference map power from the 
Sachs-Wolfe 
plateau is independent of $\ell$, apart from a logarithmic correction.  This 
is the $\ell$-dependence we expect for the Sachs-Wolfe plateau for the 
anisotropy power spectrum $\Cl$ from a strongly blue tilted primordial 
spectrum, with scalar index $n_S = 3$, as we predicted above based on 
\eq(\ref{dmpsfssmallt2}).  Comparing \eqs(\ref{cldeffs}) and 
(\ref{dmpsfssmallt}) for the power spectra of the anisotropies and of the 
difference map, and recalling the expression \eq(\ref{tdeffs}) for the 
transfer function, we see that the ``monopole'' contribution to the spectrum 
$D^{\eta\eta'}\!(\bl)$ (the part proportional to $A_1$) is proportional to 
the {\em dipole\/} contribution to the spectrum $C(\bl,\eta)$ (the part 
proportional to $A_2$).

   Finally, we note that we can evaluate \eq (\ref{corfncfs}) for the 
correlation function analytically for all $\del\xls$ for the case of a 
delta-source in $k$-space, $\PR(k) = \PR\delta(k - \tl k)$.  Such a source 
will be very helpful in understanding the temporal behaviour of the 
normalized correlation function $\bar C^{\eta\eta'}\!(\bl)$ at late times.  
The result for such a source is
\beq
\bar C^{\eta\eta'}\!(\bl) = \ld\{\begin{array}{ll}
    \cos\ld[\tl k_x(\ell)\delta\xls\rd]&\textrm{if $\ell \le \tl k\xls$},\\
    0                                  &\textrm{if $\ell >   \tl k\xls$},
    \end{array}\rd.
\label{exactcorfnc}
\eeq
where
\beq
\tl k_x(\ell) \equiv \sqrt{\tl k^2 - \fr{\ell^2}{\xls^2}}
\eeq
is the line-of-sight component of the source mode $\tl k$ corresponding 
to the observed scale $\bl$.  This result tells us that the normalized 
correlation function is initially (at $\delta\eta = 0$) unity, as 
expected, and subsequently oscillatory in $\delta\eta$, with positive 
correlations alternating with anticorrelations, and each scale $\ell$ 
oscillating at a different rate.  The largest angular scales (smallest 
$\ell$) reach anticorrelation first, followed by smaller scales.  The peak 
scale, $\ell = \tl k\xls$, never becomes anticorrelated.  This behaviour 
can be easily understood with the assistance of \fig\ref{anticorfig}, by 
noting that at the peak $\ell$ scale we have $\tl k_x(\ell) = 0$, so that 
the modes $\bk$ which contribute to the peak $\ell$ scale are parallel to 
the LSS and hence cannot produce anticorrelations.  As $\ell$ decreases, 
$\tl k_x(\ell)$ increases, \ie\ $\bk$ contains an increasing component 
parallel to the line of sight, so the first anticorrelations occur earlier 
and earlier.  If we consider sources at different scales $\tl k$, 
\eq(\ref{exactcorfnc}) tells us that the first anticorrelations occur 
earlier for smaller scales (larger $\tl k$), as expected.

\begin{figure}[ht]
\includegraphics[width=\columnwidth]{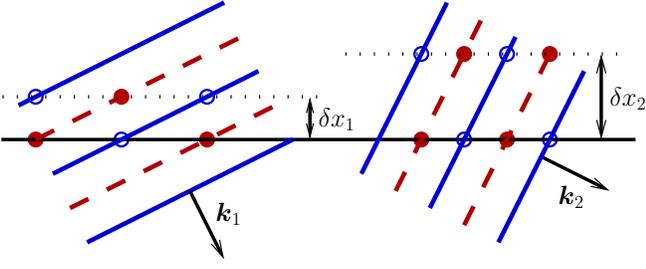}
\caption{Two source modes, $\bk_1$ and $\bk_2$, with the same wavenumber 
$\tl k$.  Diagonal lines indicate troughs (solid) and crests (dashed).  
The observer is towards the bottom.  The solid horizontal line indicates 
the position of the LSS at initial time $\eta$.  The dotted horizontal 
lines indicate the position of the LSS at the first later time 
that produces perfect anticorrelations with the initial time, so 
that hot spots (solid circles) line up with cold spots (open circles) 
and vice versa.  
Mode $\bk_1$, which corresponds to a smaller observed $\ell$, reaches 
anticorrelation before mode $\bk_2$ (since $\del x_1 < \del x_2$).  A mode 
$\bk$ parallel to the LSS would never reach anticorrelation, while a mode 
parallel to the line of sight would correspond to $\ell = 0$ (in the flat 
sky approximation).}
\label{anticorfig}
\end{figure}

\subsubsection{Origin of the difference map power}

   As we mentioned above, the power spectrum of the difference map, 
$D_\ell^{\eta\eta'}$, contains two distinct contributions: the loss of 
correlations and the change in variance between the two times of 
observation.  To make this explicit, and to determine which contribution 
is more important, we can use \eqs(\ref{dmpsdef}) and 
(\ref{normcorfnc}) to write
\bea
D_\ell^{\eta\eta'} &=& \ld(\sqrt{\Cl(\eta')} -\sqrt{\Cl(\eta)}\rd)^2\nonumber\\
   && +\ 2\sqrt{\Cl(\eta)\Cl(\eta')}\ld(1 - \bar{C}_\ell^{\eta\eta'}\rd)\\
   &=& \Cl(\eta)\ld[\oo{4}\ld(\fr{\del\Cl}{\Cl(\eta)}\rd)^2
      + 2\ld(1 - \bar{C}_\ell^{\eta\eta'}\rd)\rd]\nonumber\\
   && +\ \O\ld(\fr{\del\eta}{\eta_{\rm LS}}\rd)^3.
\label{diffmapdecomp}
\eea
The first line above is exact, while in the second we have 
dropped higher order terms in
\beq
\fr{\del\Cl}{\Cl(\eta)} \sim \fr{\del\eta}{\eta_{\rm LS}}
\label{delcldeleta}
\eeq
[recall \eq(\ref{deltaclexpl})].  With a calculation similar to that 
leading to \eq(\ref{dmpsfssmallt}), it is straightforward to show that, 
for short time intervals ($\del\eta/\eta_{\rm LS} \ll 1)$, we have 
$1 - \bar{C}_\ell^{\eta\eta'} \propto (\del\eta/\eta_{\rm LS})^2$, so that 
the two terms in square brackets in \eq(\ref{diffmapdecomp}) are of the 
same order in $\del\eta/\eta_{\rm LS}$.

   The first term in square brackets in the expression (\ref{diffmapdecomp}) 
is due entirely to the change in variance $\del\Cl$, while the second term 
is due solely to the loss of correlations between $\alm(\eta)$ and 
$\alm(\eta')$ [recall \eq(\ref{diffmappsalt})].  But from 
\eq(\ref{dmpsfssmallt2}) we have
\beq
D_\ell^{\eta\eta'} \sim \ld(\fr{\del\eta}{\eta_{\rm LS}}\rd)^2\ell^2\Cl(\eta).
\label{dmpsorder}
\eeq
This expression dominates the change in variance contribution to 
\eq(\ref{diffmapdecomp}) by a factor $\ell^2$.  
Therefore, for all but the very largest angular scales (smallest $\ell$), 
the second term in the brackets in (\ref{diffmapdecomp}) must dominate 
over the first, and so the power spectrum $D_\ell^{\eta\eta'}$ is dominated 
by the loss in correlations.  This can be confirmed by a direct computation 
in the flat sky approximation, which gives
\beq
2C(\bl,\eta)\ld(1 - \bar C^{\eta\eta'}\!(\bl)\rd) = D^{\eta\eta'}\!(\bl),
\eeq
at lowest order in $\del\eta/\eta_{\rm LS}$.  This means that the flat sky 
approximation to $D_\ell^{\eta\eta'}$ captures {\em only\/} the (dominant) 
contribution due to loss of correlations.  This is not surprizing: because 
of the scaling relation (\ref{clscal}), the flat sky difference map 
defined in \eq(\ref{diffmapdeffs}) is closely related to the 
``normalized'' difference map defined in \eq(\ref{diffmapdefalt}).

   One further contribution to the difference map arises if we consider 
the absolute temperature anisotropies $\del T$ rather than the relative 
quantity $\del T/T$, where $T$ is the mean temperature.  Recall from 
Section~\ref{sec:analps} that, if we consider the absolute spectrum 
$T^2\Cl$ instead of the relative quantity $\Cl$, then the difference 
$\del(T^2\Cl) \equiv T^2(\eta')\Cl(\eta') - T^2(\eta)\Cl(\eta)$ 
receives an extra contribution due to the expansion redshift.  In that 
case we showed that the extra contribution is of the same order as the 
geometrical scaling part [recall \eq(\ref{deltat2cl})].

   We can now repeat this calculation for the difference map power 
spectrum.  The difference map in absolute temperature units is
\beq
\del(T\alm) = T\ld[-\fr{\del\eta}{(aH)^{-1}}\alm + \del\alm\rd]
\label{absdiffmap}
\eeq
at lowest order in $\del\eta/(aH)^{-1}$, where we have used $\dot T = -HT$.  
Therefore the corresponding power spectrum becomes
\bea
\bra\del(T\alm)\del(T\alm)^\ast\ket&& = T^2\ld[D_\ell^{\eta\eta'} + 
   \ld(\fr{\del\eta}{(aH)^{-1}}\rd)^2\Cl\rd.\nonumber\\
   &&+ \fr{\del\eta}{(aH)^{-1}}\ld(D_\ell^{\eta\eta'} - \del\Cl\rd)\Bigg],
\eea
where we have used the expressions (\ref{diffmapps}), 
(\ref{dmpsdef}), and (\ref{corfncdef}).  Next, retaining only terms to 
lowest order in $\del\eta/(aH)^{-1} \sim \del\eta/\eta_{\rm LS}$, and 
using \eq(\ref{delcldeleta}), we have
\beq
\bra\del(T\alm)\del(T\alm)^\ast\ket = T^2\ld[D_\ell^{\eta\eta'} + 
   \O\ld(\fr{\del\eta}{\eta_{\rm LS}}\rd)^2\Cl\rd].
\label{absdmps}
\eeq
But then \eq(\ref{dmpsorder}) tells us that the first term on the right-hand 
side of \eq(\ref{absdmps}) dominates for all but the very largest angular 
scales [just as we argued above for \eq(\ref{diffmapdecomp})], and so
\beq
\bra\del(T\alm)\del(T\alm)^\ast\ket \simeq T^2D_\ell^{\eta\eta'}.
\eeq
In other words, the part of the power spectrum for the absolute difference 
map $\del(T\alm)$ which is due to the expansion redshift is subdominant.  
Thus, contrary to the case with $\del\Cl$, it is irrelevant for the 
difference map whether we consider absolute or relative temperature 
differences (apart from on the very largest scales).

   In hindsight this result could have been anticipated directly from 
\eq(\ref{absdiffmap}), since we expect that the change $\del\alm$, 
corresponding to the time interval $\del\eta$, should be
\beq
\del\alm \sim \fr{\del\eta}{\eta_{\rm LS}}\ell\alm,
\label{delalmcrude}
\eeq
so that the first term on the right-hand side of \eq(\ref{absdiffmap}), 
which is due to the expansion redshift, is subdominant on all but the 
largest scales.  Intuitively, the change in $\alm$ due to a change in 
observation time $\del\eta$ grows as the wavelength of the source modes 
decreases (for constant mode amplitude), since the corresponding 
increase in radius of the LSS is a larger fraction of a shorter wavelength 
mode.  On the other hand, the change in $T\alm$ due to the expansion 
redshift is independent of scale $\ell$.

   Similarly, the contribution to the difference map due to loss of 
correlations, which is described crudely by \eq(\ref{delalmcrude}), is 
expected to dominate over the contribution due to changing variance $\Cl$, 
which is roughly independent of $\ell$, as we showed rigorously above.


\subsection{Time evolution from CAMB}
\subsubsection{Power spectrum and correlation function}

   We have computed the correlation function $\Cl^{\eta\eta'}$ from 
\eq(\ref{corfnc}) and the difference map power spectrum $D_\ell^{\eta\eta'}$ 
from \eq(\ref{dmpsdef}) numerically using our modified version of 
\camb\ to extract $T(k,\ell,\rls)$ at different $\rls$ (as outlined 
in Section~\ref{sec:clcamb}), using the cosmological parameters of our 
fiducial $\Lambda$CDM model.  In \fig\ref{difpscambanal} we display 
$D_\ell^{\eta\eta'}$ for the times $\eta$ and $\eta'$ corresponding to 
today, $a_{\rm obs} = 1$, and future times when $a'_{\rm obs} = 
1 + \delta a$, for the cases $\delta a = 10^{-4}$, $0.001$, $0.01$, and 
$0.1$.  For small increments $\delta a$ these curves exhibit precisely the 
quadratic scaling $D_\ell^{\eta\eta'} \propto (\delta \eta)^2$ that we 
predicted in \eq(\ref{difftdep}), and the slope of $D_\ell^{\eta\eta'}$ 
for small $\ell$ matches our analytical prediction for the Sachs-Wolfe 
plateau, \eq(\ref{corfncsw}).  For large increments $\delta a$ the 
difference map power spectrum approaches the sum of the individual power 
spectra as the correlation function decays to zero, as we expect according 
to \eq(\ref{dmpsdef}).  Generally, these curves exhibit the heavily 
blue-tilted form we predicted in the previous subsection, due to the 
more rapid loss of correlations on smaller angular scales.

\begin{figure}[ht]
\includegraphics[width=\columnwidth]{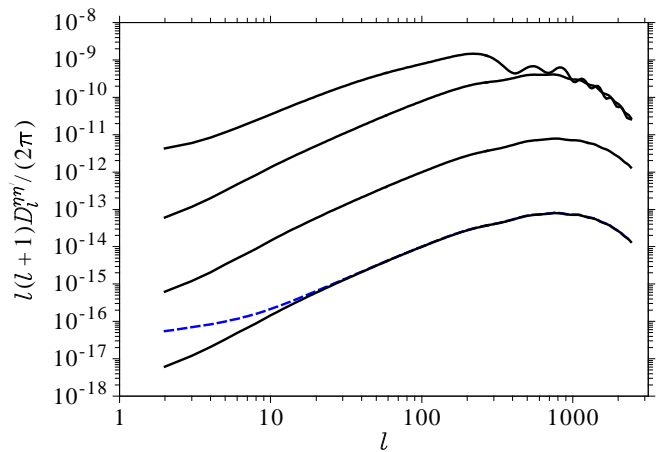}
\caption{Difference map angular power spectrum $D_\ell^{\eta\eta'}$ 
calculated from our modified version of \camb\ (solid lines) for the two 
times 
corresponding to $a_{\rm obs} = 1$ and $a'_{\rm obs} = 1 + \delta a$, for 
the cases (top to bottom) $\delta a = 0.1$, $0.01$, $0.001$, and $10^{-4}$.  
Also shown is the analytical result (dashed curve) calculated using 
\eq(\ref{dmpsfssmallt2}) for $\delta a = 10^{-4}$.}
\label{difpscambanal}
\end{figure}

   Also shown in \fig\ref{difpscambanal} is the curve calculated from 
the flat-sky analytical expression, \eq(\ref{dmpsfssmallt2}), for the 
case $\delta a = 10^{-4}$.  This curve coincides extremely well with the 
numerical result for $\ell \agt 20$.  The departures at large scales are 
due to two factors.  First, the flat sky approximation is poor at those 
scales.  Second, \eq(\ref{dmpsfssmallt2}) was derived under the assumption 
that all anisotropies were primary, which is not the case for the ISW 
contribution.

   In Fig.~\ref{corfncfig} we plot the normalized correlation function 
$\bar C_{\ell}^{\eta\eta'}$, calculated using our modified version of 
\camb\ for our 
fiducial $\Lambda$CDM model, between the set of $a_{\ell m}$s observed at 
$\eta$ and $\eta'$.  Here, $\eta$ corresponds to an observation of the CMB 
sky today at $a_{\rm obs} = 1$ and $\eta'$ to an observation at 
$a'_{\rm obs} = 1 + \delta a$, where we illustrate the cases $\delta a = 
0.001$, $0.01$, $0.03$ and $0.1$.  For the smallest interval $\delta a$, 
we find very strong correlation between the two sky maps, as expected.  
The correlations fall off as $\delta a$ increases, with the sky maps 
becoming somewhat anticorrelated for intermediate intervals before 
$\bar C_{\ell}^{\eta\eta'}$ decays to zero at the largest intervals.

\begin{figure}[ht]
\includegraphics[width=\columnwidth]{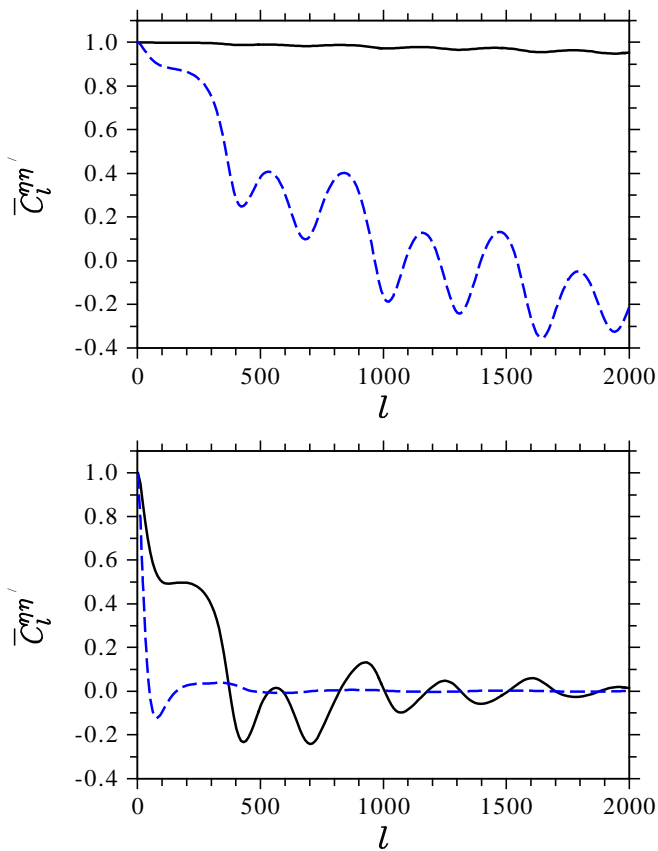}
\caption{Normalized correlation function 
$\bar C_\ell^{\eta\eta'}$ between the CMB sky observed today ($a = 1$) and 
at $a = 1 + \delta a$, calculated with our modified version of \camb, for 
$\delta a = 0.001$ (top panel, solid curve), $0.01$ (top panel, dashed 
curve), $0.03$ (bottom panel, solid), and $0.1$ (bottom panel, dashed).  
Anticorrelations are seen 
to develop as $\delta a $ increases, before the correlation function decays 
to zero for large $\delta a$.}
\label{corfncfig}
\end{figure}

   The general features of the correlation function can be understood 
by considering the detailed arguments presented in the previous subsection.  
For $\delta a \ll 1$ the increase in the LSS radius corresponding to the 
interval $\delta a$ is $\delta\rls = H_{0}^{-1}\delta a$.  For the case of 
$\delta a = 0.01$, using $H_0 = 73$ ${\rm km}\,{\rm s}^{-1}{\rm Mpc}^{-1}$, 
we find $\delta\rls = 40.9$ Mpc, corresponding to a comoving wavenumber 
$k = 0.15$ ${\rm Mpc^{-1}}$.  This wavenumber is much larger than the 
wavenumber of the first acoustic peak, given by $k = \pi/s_{\rm LS} = 0.021$ 
${\rm Mpc^{-1}}$, where $s_{\rm LS} \simeq 150$ Mpc is the sound horizon 
at last scattering.  Hence, for this $\delta a$, we are essentially 
sampling the {\em same\/} set of inhomogeneities which give rise to the 
first acoustic peak at both times, and so we expect fluctuations to be 
correlated on these scales.  Indeed, we see from \fig\ref{corfncfig} for 
$\delta a = 0.01$ that $\bar C_{\ell}^{\eta\eta'} \simeq 0.9$ for the first 
acoustic peak scale, $\ell \simeq 220$.  Extending this argument, we 
expect that $\bar C_{\ell}^{\eta\eta'} \ra 1$ as $\ell \ra 0$ for fixed 
$\delta a$, as the largest scale (smallest $k$) features should be most 
correlated, and of course we similarly expect $\bar C_{\ell}^{\eta\eta'} 
\ra 1$ as $\delta a \ra 0$ for fixed $\ell$.

   The presence of anticorrelations was discussed in 
Section~\ref{sec:analdiffmap}, where we derived the behaviour of the 
correlation function in the flat sky approximation for the case of a 
delta-source at wavenumber $k = \tl k$.  The result, \eq(\ref{exactcorfnc}), 
exhibited oscillating positive and negative correlations, with the first 
anticorrelations occuring earlier for smaller scales (larger $\tl k$), as 
we have confirmed here for a realistic spectrum using \camb.  
\eq(\ref{exactcorfnc}) also described 
anticorrelations occuring earlier for {\em smaller\/} $\ell$, with 
$\tl k$ fixed.  This behaviour is not visible in the actual correlation 
function plotted in \fig\ref{corfncfig}, since the real primordial power 
spectrum is far from being a delta-source.  If we consider a small subset of 
$k$ modes, \eg\ those corresponding to the fourth acoustic peak scale, 
then some of those modes will be aligned nearly parallel to our line of 
sight and hence produce early anticorrelations at small $\ell$ for some 
$\delta a$ (recall \fig\ref{anticorfig}).  However, there are many more 
modes due to power at smaller $k$ that are still tightly correlated at 
the same $\delta a$ and hence result in $\bar C_{\ell}^{\eta\eta'} \simeq 1$ 
for small $\ell$.


\subsubsection{Sky maps}

   Assuming Gaussianity, generating a {\em single\/} realization of a set of 
$a_{\ell m}$s usually involves drawing each $a_{\ell m}(\eta)$ 
independently from a Gaussian distribution with variance $C_{\ell}(\eta)$.  
With the correlation function $C_\ell^{\eta\eta'}$, we have a measure of 
the degree of correlation between $a_{\ell m}$s at two different times.  
Hence, given a set of $a_{\ell m}$s at the first time, the variance of 
the distribution at each time, and the correlation between them, one can 
generate a realization of a second set of $a_{\ell m}$s at some later time. 

   Formally, we draw the new set of $a_{\ell m}$s from the likelihood function
\begin{equation}
P(X_i)=\frac{1}{2\pi|C|^{1/2}} \exp \left(-\frac{1}{2} X_{i}^{T} C^{-1} X_i 
\right)\,, 
\end{equation}
with
\begin{equation}
C = \left(\begin{matrix} C_{\ell}^{\eta \eta} & \Cl^{\eta\eta^{\prime}}  
\cr  \Cl^{\eta\eta^{\prime}} & C_{\ell}^{\eta^{\prime} \eta^{\prime}} 
\end{matrix}\right)\,,
\end{equation}
where $X_i$ is a random 2-vector containing each $a_{\ell m}$ coefficient 
at $\eta$ and $\eta^\prime$, and we have relabeled the variance of the 
$a_{\ell m}(\eta)$ distribution at each time by $C_{\ell}^{\eta \eta}$ 
and $C_{\ell}^{\eta^{\prime} \eta^{\prime}}$.

   We illustrate the likelihood function in Fig.~\ref{fig:alm} for the 
$a_{2m}$ and $a_{5m}$ modes, where $\eta$ corresponds to an observation at 
$a_{\rm obs} = 1$ and $\eta^\prime$ to $a_{\rm obs} = 2.0$.  The $a_{2m}$ 
coefficients are more tightly correlated than their $a_{5m}$ counterparts, 
since for such a large $\delta a$ the correlation rapidly falls off as 
$\ell$ increases.  It is also noticeable that the contours of the 
likelihood are slightly elongated vertically, due to the increased variance 
of $a_{\ell m}(\eta^\prime)$ on large scales resulting from the increasing 
ISW effect. 

\begin{figure}[ht]
\includegraphics[width=\columnwidth]{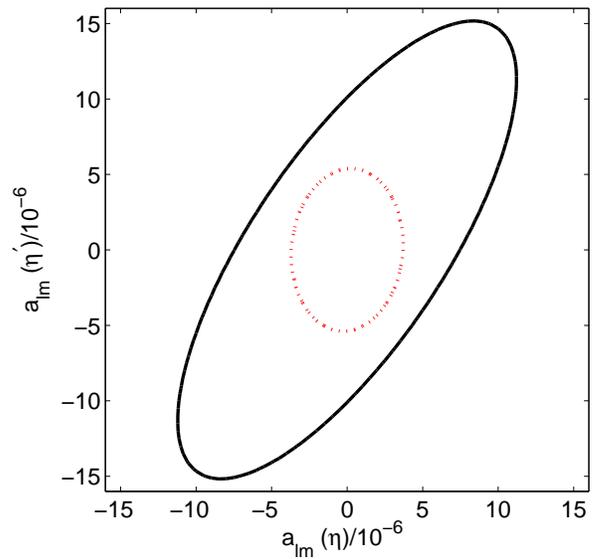}
\caption{\label{fig:alm} Distribution from which the $a_{\ell m}$s are 
drawn. Here, $\eta$ corresponds to $a_{\rm obs}=1$ and $\eta^\prime$ 
to $a_{\rm obs}=2.0$, and contours show the $2 \sigma$ error ellipse.  
The distribution of $a_{2m}$ is shown by the solid contour, which has a 
correlated set of variances $C_{2} (\eta)$ and $C_{2} (\eta^\prime)$. The 
distribution of $a_{5m}$, shown by the dotted contour, has a smaller 
variance at both times and these are much less correlated.}
\end{figure}

   Therefore, our method of generating CMB sky maps involves firstly 
generating a random realization at some initial time, and then generating 
all subsequent realizations by mapping the $a_{\ell m}$s using the 
correlation function. For large $\delta a$, where the $a_{\ell m}$s 
are uncorrelated, we are essentially selecting a completely new set of 
coefficients.  For small $\delta a$, $\bar C_{\ell}^{\eta\eta'}$ approaches 
unity and the $a_{\ell m}$s map trivially according to $a_{\ell m}(\eta) 
\rightarrow a_{\ell m}(\eta^\prime)$.  At some intermediate intervals, 
anticorrelation favours a reversal of sign of the $a_{\ell m}$s, i.e.\ 
hot spots are mapped to cold spots and vice versa.

   We generate maps using the HEALPix code \footnote{Information on 
HEALPix is available at \href{http://healpix.jpl.nasa.gov}
{http://healpix.jpl.nasa.gov}.} with $n_{\rm side}=512$, 
corresponding to a pixel resolution of $6.87$ arcmin. We present a series 
of these maps in Fig.~\ref{fig:maps}, plotting the fractional temperature 
fluctuation $\delta T_{i}/T$ at each pixel $i$. For presentational 
clarity we show a patch of sky covering $\sim 1000$ square degrees, 
and use modes up to $\ell_{\rm max}=1000$. We generate the first map 
at $a_{\rm obs}=1$, and show subsequent maps at $a_{\rm obs}=1+\delta a$, 
where $\delta a=0.001$, $0.01$, $0.1$, and $1.0$. We also show the 
difference map for each observation relative to $a_{\rm obs}=1$.  We 
have checked that the power spectra reconstructed from our simulated 
sky maps agree with the intended spectra to within sample variance.  
Note that for the sky maps in Fig.~\ref{fig:maps} we did not use the 
actual WMAP data for the present time; rather, we simply generated 
a random initial map according to the required $\Cl$ spectrum.

\begin{figure*}
\includegraphics[width=0.8\textwidth]{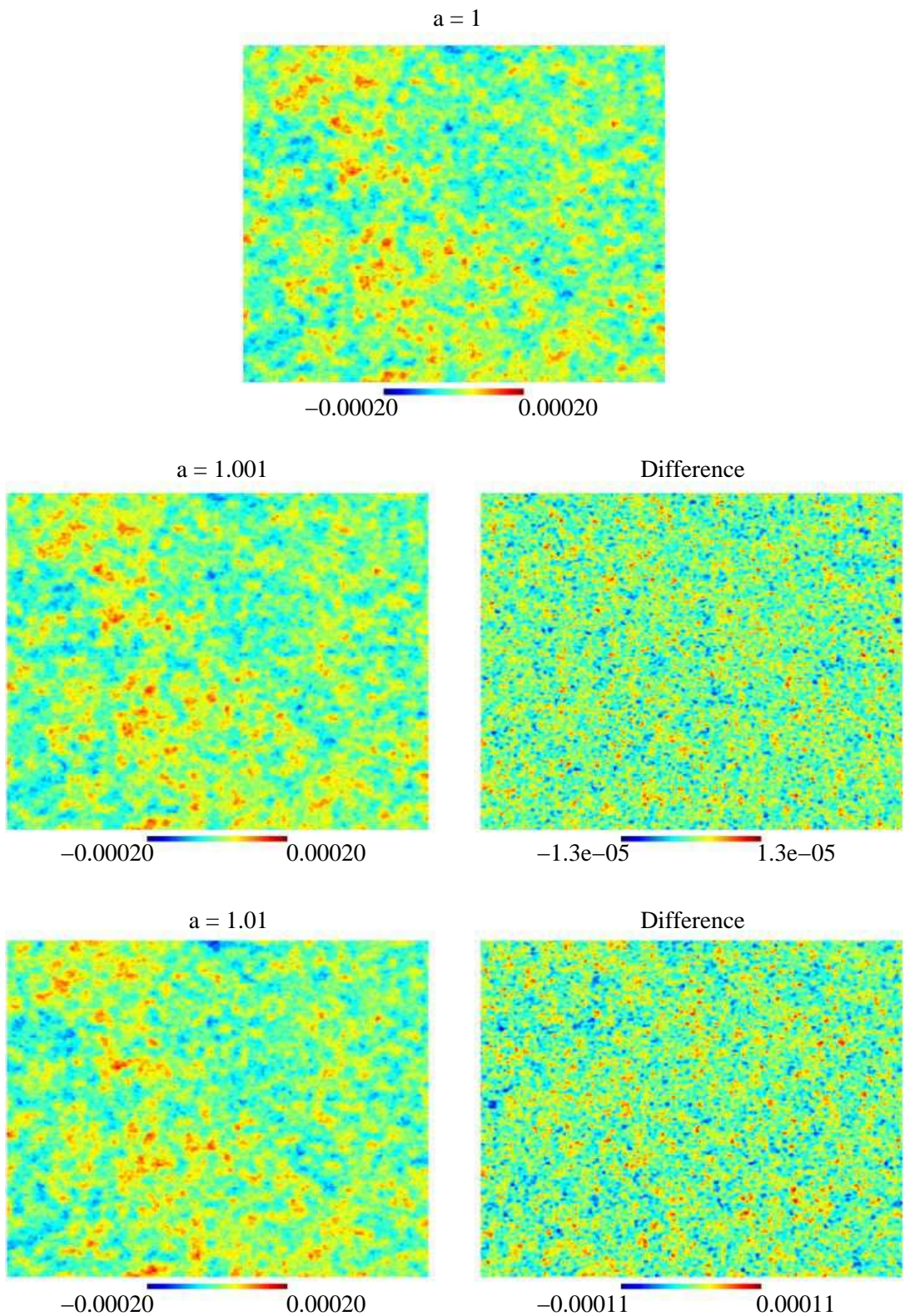}
\caption{Simulated map realizations (top and left panels) 
and difference map (relative to $a_{\rm obs}=1$) (right panels) for 
$\delta a=10^{-3}$ (middle panels, corresponding to a $13$ Myr interval) 
and $10^{-2}$ (bottom panels, $130$ Myr).  Note the vastly different 
power scales between the sky and 
difference maps.  The maps presented here are for a patch of sky covering 
$\sim 1000$ square degrees.  High resolution version available at 
\href{http://www.astro.ubc.ca/people/scott/future.html}
{http://www.astro.ubc.ca/people/scott/future.html}.}
\label{fig:maps}
\end{figure*}

\addtocounter{figure}{-1}
\begin{figure*}
\includegraphics[width=0.8\textwidth]{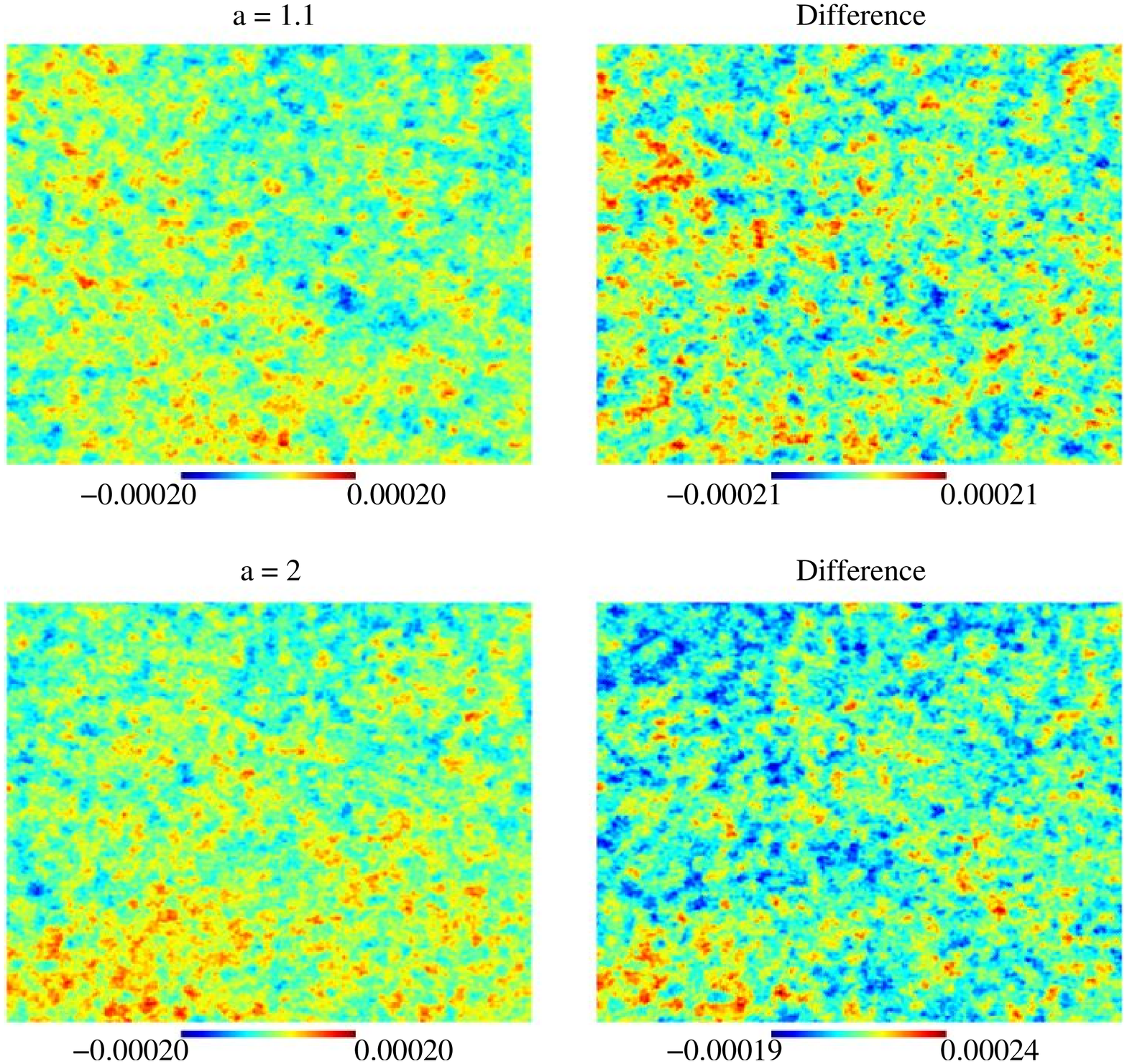}
\caption{(Continued.) Simulated map realizations (left 
panels) and difference (right) for $\delta a=0.1$ (top) and $1.0$ (bottom).  
High resolution version available at 
\href{http://www.astro.ubc.ca/people/scott/future.html}
{http://www.astro.ubc.ca/people/scott/future.html}.}
\label{fig:maps2}
\end{figure*}

   Visually, the $\delta a=0.001$ map is extremely similar to the initial 
map. The variance of the map, given by
\beq
\left\langle \left( \frac{\delta T}{T} \right)^{2} \right\rangle_{\rm map} = 
\frac{1}{N_{\rm pix}} \sum_i^{N_{\rm pix}} \left( \frac{\delta T_i}{T}
\right)^2\,, 
\eeq
is over four orders of magnitude higher than the difference map variance. 
For $\delta a=0.01$, the primary temperature fluctuations have a variance 
around two orders of magnitude more than the difference map, and changes 
in small scale structure  (from the initial map) are clearly apparent. 

   For $\delta a=0.1$ and $1.0$, the variance of the difference is actually 
larger then the temperature fluctuations at that time, and acoustic scale 
structures are visible in the difference. This is understandable from our 
discussion of the correlation function---at these times the correlation on 
all but the very largest scales has dropped to zero, so that the variance 
of the difference approaches the sum of the initial and final map variances 
[recall \eq(\ref{dmpsdef})].

   Finally, in \fig\ref{ultimatemap} we present a simulated sky map for 
the asymptotic future.  This map clearly differs from today's map, with 
the dramatic increase in large scale power due to the ISW effect readily 
apparent.  For this map, we derived the initial $\alm$ coefficients up 
to $\ell = 20$ from the WMAP Internal Linear Combination map 
\cite{hinshaw06} \footnote{We also employed the LAMBDA archive 
\href{http://lambda.gsfc.nasa.gov/}{http://lambda.gsfc.nasa.gov/}.} 
(for $\ell > 20$ we generated random initial modes instead of using the real 
data, since the normalized correlation is negligible on those scales at these 
very late times).

\begin{figure*}
\includegraphics[width=0.85\textwidth]{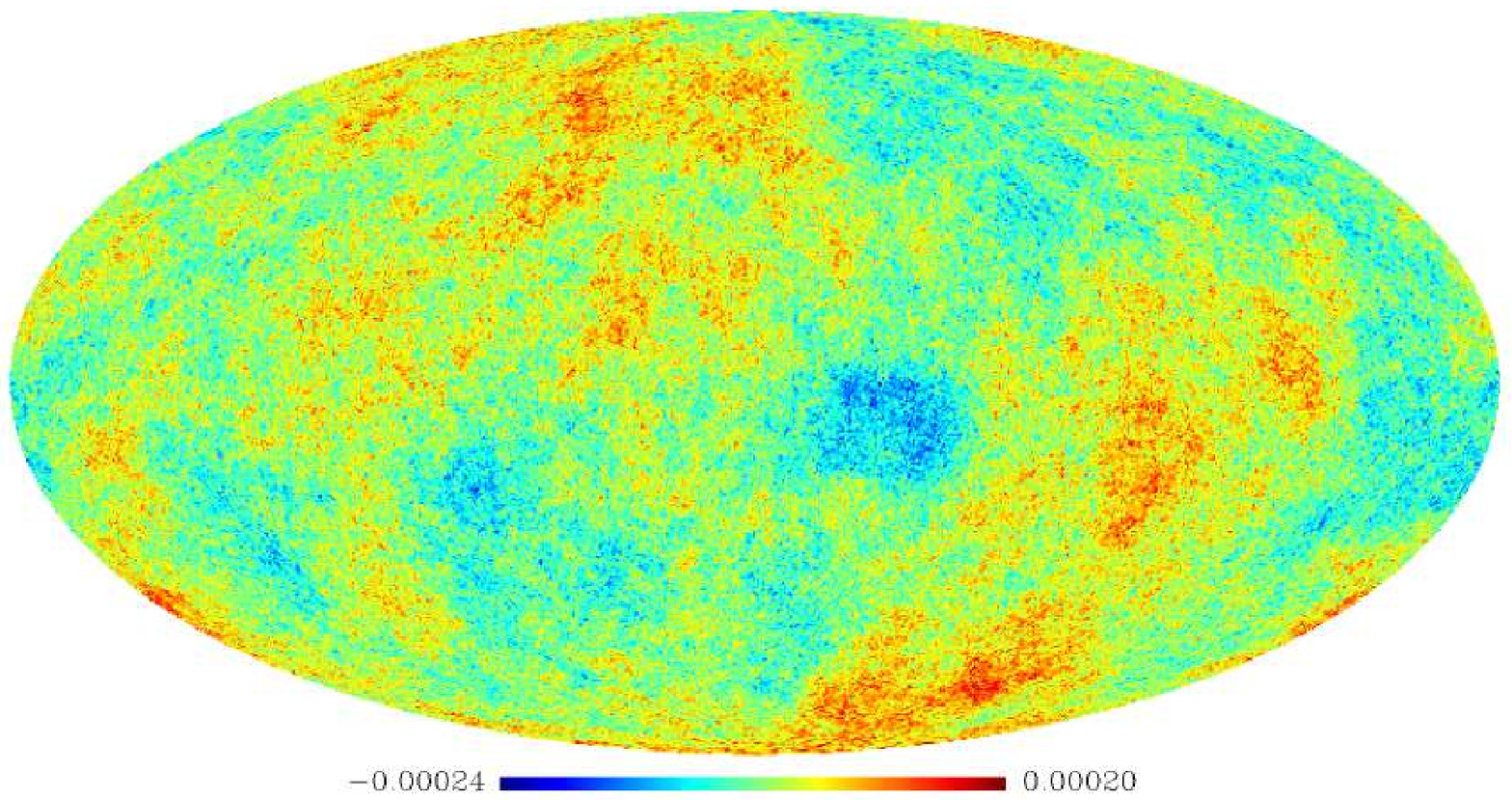}
\caption{Simulated map realization for the asymptotic future.  High 
resolution version available at 
\href{http://www.astro.ubc.ca/people/scott/future.html}
{http://www.astro.ubc.ca/people/scott/future.html}.}
\label{ultimatemap}
\end{figure*}

  High resolution versions of these sky maps, together with animations 
illustrating the evolution of the CMB sky maps, $\Cl$ spectra, and 
correlation functions, are available at 
\href{http://www.astro.ubc.ca/people/scott/future.html}
{http://www.astro.ubc.ca/people/scott/future.html}.


\section{Discussion}
\label{sec:discuss}

   We have systematically described the temporal evolution of the CMB, 
beginning with the mean temperature and dipole, and then moving to 
the anisotropy power spectrum.  We found that the evolution of the 
spectrum is described at all but the largest angular scales by a simple 
scaling relation.  At large scales the ISW contribution grows to 
dominate even the first acoustic peak at late times.  The extra optical 
depth due to reionization is negligible into the future.

   We have introduced a correlation function between the CMB sky maps 
at different times which quantitatively encodes the intuitive notion that 
for small enough observation time intervals $\del\eta$ and for source modes 
with small enough wavenumber $k$, the anisotropies observed at the two 
times should be very similar.  Closely related is the power spectrum of 
the difference map $D_\ell^{\eta\eta'}$.  We showed that the difference 
$\del\Cl$ scales like $\del\eta$ for small intervals, while 
$D_\ell^{\eta\eta'}$ scales like $(\del\eta)^2$.  The sensitivity of 
$D_\ell^{\eta\eta'}$ to changes in the sky maps is dominated by the 
loss of correlations at small angular scales, and the contributions from 
the change in variance 
$\Cl$, as well as the change due to expansion redshift (if we consider 
absolute quantities) are subdominant.  All of our numerical 
results were independently confirmed analytically, and the validity of 
the necessary analytical approximations was elucidated by the numerics.

   The quantities we described in this work will be crucial to 
answering the question of the experimental detectability of a change 
in the CMB, or, more precisely, the question ``how long must we wait to 
be able to confidently observe a change?''  While the different time 
interval scalings 
we found for $\del\Cl$ and $D_\ell^{\eta\eta'}$ might suggest that 
attempting to measure $\del\Cl$ would be much more favourable for small 
$\del\eta$, the situation is more subtle.  In a separate paper \cite{mzs07} 
we quantify the detectability of changes in the CMB.

   On sufficiently long time scales the CMB evolution we have 
described will be obvious, and hence it is natural to ask what cosmological 
information a measurement of such changes might eventually provide 
to future cosmologists.  
A measurement of the cooling rate of the mean temperature $T$ would 
provide an independent and novel determination of the local 
Hubble rate $H_0$, through the relation $\dot T = -HT$.  As far as the 
primary temperature anisotropies are concerned, according to 
\eq(\ref{dmpsfssmallt2}) the shape of the difference map spectrum 
$D_\ell^{\eta\eta'}$ is determined by the $\Cl$ spectrum [and 
the primordial spectrum through the quantity $C^{(n_S + 2)}(\bl,\eta)$], 
so a measurement of the shape of $D_\ell^{\eta\eta'}$ should not provide 
much new information.  With our assumption of a spatially flat geometry, 
the amplitude of $D_\ell^{\eta\eta'}$ is determined by the ratio 
$\delta\xls/\xls = \delta t/(a_0\xls)$.  Thus a measurement of the 
amplitude would directly fix the LSS radius $\xls$ and hence provide 
an independent constraint on the parameters $\Omega_m$, $\Omega_\Lam$, 
and $H_0$.  (This determination of $\xls$ is replaced by a determination 
of the angular diameter distance for spatially curved models.)  The 
radius of the LSS is currently fixed, through observations of the accoustic 
angular scale, only up to the uncertainty in the matter content at last 
scattering.  At very late times much additional information will of course 
become available as new modes become visible on the growing LSS.

   We have focussed entirely on primordial anisotropies here.  There 
are additional issues which arise when one considers secondary 
anisotropies, like gravitational lensing and Sunyaev-Zel'dovich effects, 
as well as time-dependent foregrounds of course.  Such considerations 
depend much more heavily on the less well understood non-linear 
scales of structure, and so we leave this for others to pursue.  We 
expect that there is plenty of time to pursue these ideas before any 
of these variations would be detectable.

   When this work was nearly complete, a related study appeared by Lange 
and Page \cite{lp07}.  Those authors calculated future $\Cl$ spectra 
using \camb.  They also defined a correlation fuction equivalent to 
our \eq(\ref{corfnc}), calculated it using \camb, and generated 
simulated future sky maps.  While they made no attempt at an analytical 
description of the evolution, their numerical results appear to agree 
with ours where they overlap.  Furthermore, they made an estimate of the 
observability of the CMB evolution, which we have deferred to 
Ref.~\cite{mzs07}.

\begin{acknowledgments}
This research was supported by the Natural Sciences and Engineering Research 
Council of Canada.  We thank Kamson Lai and Martin White for useful 
discussions, and Richard Battye for assistance with HEALPix \cite{gorski05}, 
with which some of the results in this paper have been obtained.
\end{acknowledgments}

\appendix*
\section{The flat sky approximation}
\label{flatskyapp}

   The Bessel functions appearing in the various expressions relating 
primordial fluctuations to observed CMB anisotropies severely limit the 
extent to which analytical results can be obtained.  However, a simple 
approximation scheme, based on treating a small patch of the sky (and 
hence of the spherical LSS) as flat, allows us to use ordinary plane 
wave expansions and thereby to do ``CMB without Bessel functions''.  This 
small-angle approximation is expected to be accurate up to terms of 
order $1/\ell$, so that it is entirely appropriate for describing the 
acoustic peak structure of the CMB.

   The flat sky approximation begins (see, \eg, \cite{ll00}) 
by replacing \eq(\ref{sw1}) relating the observed 
temperature anisotropies with the perturbation functions on the LSS, 
$\phi_i$, in the strong coupling/free streaming approximation, by
\beq
\fr{\delta T(\bth,\eta)}{T(\eta)} = F(\phi_i(\xls,\xls\bth,\eta_{\rm LS})).
\label{sw1fs}
\eeq
Here $\bth$ is a $2$-dimensional vector whose components represent 
the angular displacement in two orthogonal directions from the centre of 
the small patch of sky.  In the Cartesian comoving coordinate vector 
$(\xls,\xls\bth)$, the first component is parallel to, and the second 
two orthogonal to, the line of sight to the centre of the patch.  The 
coordinate value $\xls = \eta - \eta_{\rm LS}$ refers to the comoving 
distance to the LSS from the point of observation.  Analogously to 
\eq(\ref{sw2}) we can write\begin{widetext}
\beq
F(\phi_i(\xls,\xls\bth,\eta_{\rm LS})) = \phi_1(\xls,\xls\bth,\eta_{\rm LS}) +
                        \fr{\pd}{\pd\xls}\phi_2(\xls,\xls\bth,\eta_{\rm LS})
\label{sw2fs}
\eeq
for the monopole and dipole contributions.  In place of the spherical 
harmonic expansion for the temperature fluctuation, \eq(\ref{texp}), we here 
use a $2$-dimensional Fourier expansion in terms of the continuous vector 
$\bl$ which replaces $\ell$ and $m$:
\beq
\fr{\delta T(\bth,\eta)}{T(\eta)}
    = \oo{2\pi}\int d^2\bl\,a(\bl)e^{i\bth\cdot\bl}.
\label{texpfs}
\eeq

   The statistical properties of the coefficients $a(\bl)$ can be determined 
in a manner completely analogous to that used for the spherical case in 
Section~\ref{sec:form}.  Fourier expanding the perturbations $\phi_i$ 
according to
\beq
\phi_i(\xls,\xls\bth) = \oo{(2\pi)^{3/2}} \int d^2\bkp dk_x\,
                     \phi_i(\bk) e^{i\bkp\cdot\xls\bth}e^{ik_x\xls},
\label{phiexpfs}
\eeq
where $k_x$ and $\bkp$ are Cartesian components of the wavevector $\bk$ 
parallel and orthogonal to the line of sight, respectively, allows us to 
identify
\beq
\bl = \xls\bkp.
\label{blbkp}
\eeq
This tells us that $\bl$, the flat sky approximation to the spherical indices 
$\ell$ and $m$, is directly proportional to the component of the LSS 
fluctuation wavevector orthogonal to the line of sight, and that the 
relationship scales with the conformal time (or comoving distance) to the 
LSS, exactly as expected.  Since $\bkp$ is only a {\em component\/} of the 
wavevector $\bk$, \eq(\ref{blbkp}) encodes the familiar fact that the mapping 
from $k$ to $\ell$ is not one-to-one---rather, a range of $k$'s is mapped to 
a particular $\ell$ value.

   Using these expressions, we find
\beq
a(\bl,\eta) = \oo{\sqrt{2\pi}\xls^2}
   \int_{-\infty}^\infty dk_x \R(k_x,\bl/\xls) T_{\rm FS}(k,k_x) e^{ik_x\xls},
\label{almrtrfs}
\eeq
\end{widetext}where the flat sky transfer function is
\beq
T_{\rm FS}(k,k_x) \equiv A_1(k) + ik_x A_2(k),
\label{tdeffs}
\eeq
and the $A_i$ are again defined by \eq(\ref{phitransfr}).  Finally, using 
the statistical properties of $\R$ encoded in \eq(\ref{primps}), the 
equal-time correlation function of $a(\bl)$ becomes
\beq
\bra a(\bl,\eta) a^\ast(\bl\,',\eta)\ket = C(\bl,\eta)\delta^2(\bl - \bl\,'),
\label{almclfs}
\eeq
where
\beq
C(\bl,\eta) \equiv \fr{\pi}{\xls^2}\int_{-\infty}^\infty
                   dk_x\fr{\PR(k)|T(k,k_x)|^2} {(k_x^2 + \bkp^2)^{3/2}}.
\label{cldeffs}
\eeq
Again, we find that the coefficients $a(\bl)$ for different $\bl$ modes 
are uncorrelated.

   Notice that the time dependence of $C(\bl,\eta)$ is carried in the 
prefactor $1/\xls^2$ as well as in the terms containing $\bkp$ through 
\eq(\ref{blbkp}) (if $\bl$ is held constant), whereas in the spherical case, 
\eq(\ref{cldef}), the Bessel functions carry the time dependence.  Also, 
the complete absence of oscillatory functions in \eq(\ref{cldeffs}) means 
that it will be much easier to evaluate the CMB spectrum in the flat sky 
approximation than in the spherical case, both analytically and numerically.

   In particular, we can easily apply \eq(\ref{cldeffs}) to rederive the 
scaling relation \eq(\ref{clscal}).  \eq(\ref{blbkp}) tells us that $\bkp$ 
is invariant under the transformation $\xls \ra \xls'$ and $\bl \ra \bl\,' 
= \bl\xls'/\xls$.  Therefore, \eq(\ref{cldeffs}) immediately implies that
\beq
\ell'^2C(\bl\,',\eta') = \ell^2C(\bl,\eta),
\label{clscalfs}
\eeq
where $\ell \equiv |\bl|$, regardless of the form of the transfer functions 
$A_i(k)$ or of the primordial power spectrum $\PR(k)$.


\bibliography{bib}

\end{document}